\documentclass[aps,prl,showpacs,floats,twocolumn,floats,superscriptaddress,floatfix]{revtex4-1}
\usepackage{graphicx}
\usepackage{bm}
\usepackage{amsfonts}
\usepackage{color}
\usepackage{amsmath}    
\usepackage{epsfig}
\usepackage{subfigure}  
\usepackage{hyperref}   
\usepackage{bm}
\usepackage{amssymb}
\usepackage{soul}
\usepackage{xcolor}
\usepackage{tikz}
\usepackage{longtable}
\usepackage{multirow}

\begin{document}
\title{Orientation dynamics of sedimenting anisotropic particles in turbulence}
\author{Prateek Anand}
\email{prateek@jncasr.ac.in}
\affiliation{Jawaharlal Nehru Centre for Advanced Scientific Research, Bangalore 560064, India}
\author{Samriddhi Sankar Ray}
 \email{samriddhisankarray@gmail.com}
\affiliation{International Centre for Theoretical Sciences, Tata Institute of Fundamental Research, Bangalore 560089,India}
\author{Ganesh Subramanian}
 \email{sganesh@jncasr.ac.in}
\affiliation{Jawaharlal Nehru Centre for Advanced Scientific Research, Bangalore 560064, India}

\begin{abstract}

We examine the dynamics of small anisotropic particles (spheroids) sedimenting
through homogeneous isotropic turbulence using direct numerical simulations and
theory. The gravity-induced inertial torque acting on sub-Kolmogorov spheroids
leads to pronouncedly non-Gaussian orientation distributions localized
about the broadside-on\,(to gravity) orientation. Orientation distributions and
average settling velocities are obtained over a wide range of spheroid aspect
ratios, Stokes and Froude numbers. Orientational moments from the simulations
compare well with analytical predictions in the inertialess rapid-settling limit, with both exhibiting a non-monotonic dependence on spheroid aspect ratio. Deviations arise at Stokes numbers of order unity due to a spatially inhomogeneous 
particle concentration field resulting from a preferential sweeping effect; as a consequence,
the time-averaged particle settling velocities exceed the orientationally averaged estimates.

\end{abstract}

\maketitle  

Suspended inertial anisotropic particles show up in a variety of scenarios ranging from pollen dispersion to soot emission. Prominent examples in nature include ice crystals suspended in high-altitude Cirrus clouds which are a crucial element in the planetary greenhouse effect~\cite{Liou,Pandit2015}. The radiative properties of such clouds depend sensitively on the orientation distribution of ice crystals~\cite{baran}. The latter come in a variety of pristine shapes with sizes ranging from tens to thousands of microns~\cite{review}, smaller than the typical Kolmogorov scales, about a millimeter, for atmospheric turbulence. Therefore, a first step towards understanding Cirrus cloud radiation is to examine how sub-Kolmogorov anisotropic particles orient themselves while sedimenting in a turbulent flow.

The critical role of turbulence in gravitational settling has been investigated in-depth only for inertial \textit{spherical}  particles~\cite{Yau2000,monchaux2012,mehlig2016}. In this simpler scenario, relevant to the dynamics of water droplets in warm clouds, for instance, we now have a detailed understanding of the  role of turbulence in enhancing single-particle sedimentation~\cite{maxey_87,maxey_93,BecRay2014} as well as
collision~\cite{Collins2016a,Collins2016b,Wang1,Wang2,Bodenschatz,James} and coalescence~\cite{BecPRE} rates which control raindrop formation~\cite{falkovich2002,shaw2003}.

The effect of inertia for anisotropic particles is far more involved owing to additional rotational degrees of freedom~\cite{Voth2017}. Most earlier studies ignore either inertia~\cite{Pumir-NJP,VincenziPRE} (the suspended particles acting as probes for the turbulent velocity-gradient tensor~\cite{Voth2017,Meneveau2011}) or gravity~\cite{RoyPRE,Gupta}. Experiments have also largely focussed on neutrally buoyant anisotropic tracers in turbulence~\cite{Voth1,Voth2,Voth3}. Thus, gravitational settling of heavy anisotropic particles, beyond simple laminar flows under Stokesian conditions~\cite{Maxey1,Maxey2}, remains largely unexplored~\cite{Voth2017}. Recent efforts address the issue of how such particles sediment in non-trivial flows~\cite{pumirmehlig2017,siewert1,siewert2,pumir}, but the effect of gravity on rotational dynamics is not accounted for, leading
to orientation distributions that are far from being
representative. There exist efforts analyzing the motion of anisotropic particles in turbulent channel flow, the object of interest often being the particle deposition rate onto walls\,\cite{Zhao2006,Fan1995,Zhao2014,Mortensen2008,Coletti2019}. Gravity is omitted in most of these efforts; those that do include gravity again neglect its role in the rotational dynamics\,\cite{Mclaughlin2001}. In this work, using direct numerical simulations (DNSs) and theory, we characterize the distribution of particle orientations in suspensions of spheroids sedimenting in an ambient homogeneous isotropic turbulent field. Rigorously accounting for the effects of gravity on both the particle translational and rotational degrees of freedom, we find, in contrast to earlier efforts~\cite{pumirmehlig2017,siewert1,siewert2,pumir}, that the orientation distributions always peak at the broadside-on\,(to gravity) orientation. Further, although the particle settling velocities equal the orientationally averaged estimates in the rapid-settling limit, they consistently exceed the latter when effects of particle inertia become significant.
\begin{figure*}
\includegraphics[width=1.0\columnwidth]{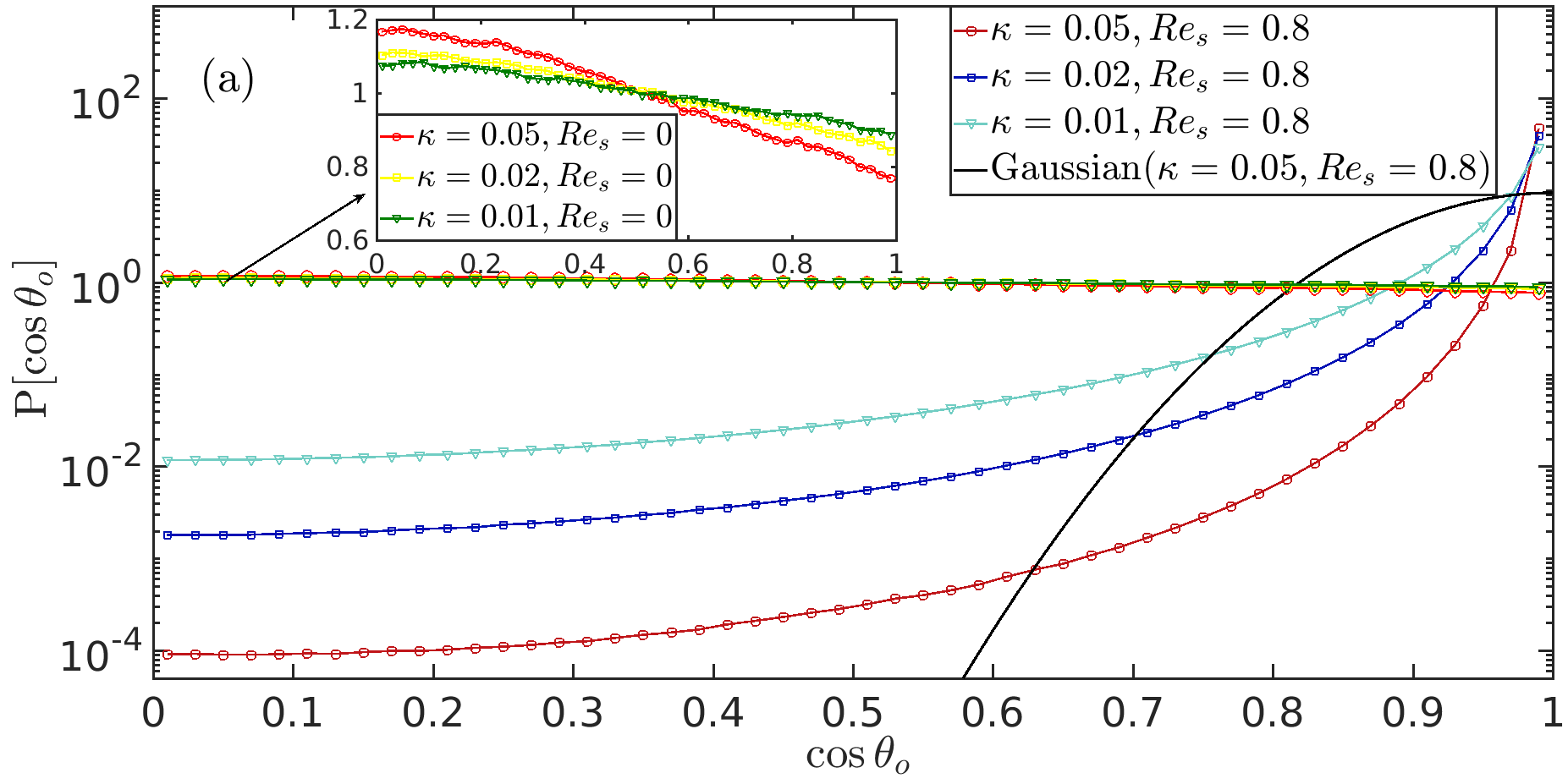}
\includegraphics[width=1.0\columnwidth]{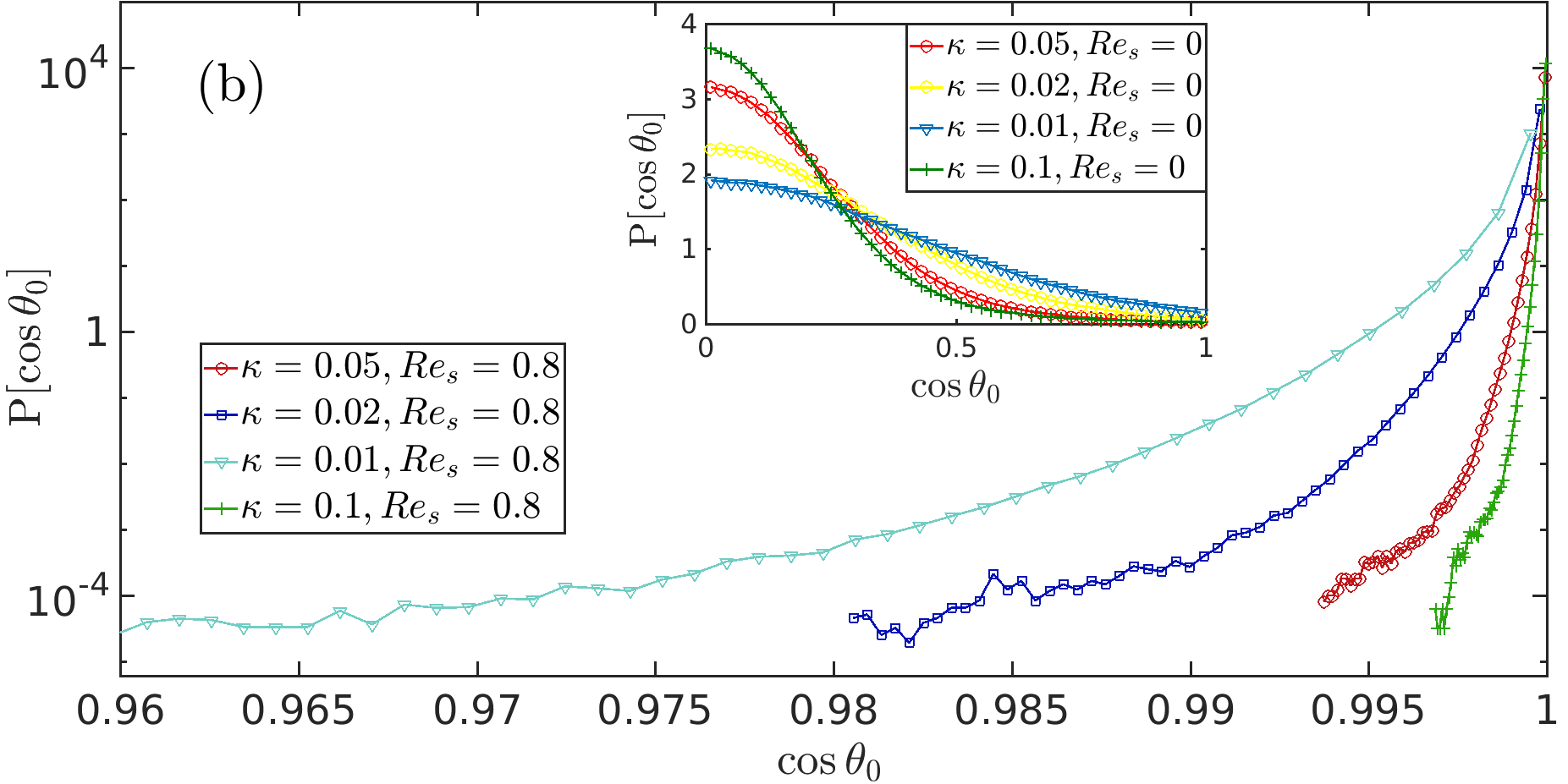}
\caption{Probability distributions of $|\bm{p}\cdot\hat{\bm g}|$ for (a) $R_\lambda = 150$
and (b) $R_\lambda = 47$ in the presence ($Re_s = 0.8$) and (inset) absence ($Re_s = 0$) of a gravity-induced torque; 
different curves correspond to different $\kappa$ (see legend). The solid black line in (a) denotes a Gaussian with the same second moment as the distribution for $\kappa =0.05$, $Re_s = 0.8$.}
\label{fig:pdf}
\end{figure*}

We perform direct numerical simulations of non-interacting spheroids sedimenting through
homogeneous isotropic turbulence with a mass loading
assumed small enough for carrier-fluid turbulence to
remain unaffected(a one-way coupled framework)\cite{SM}. The fluid velocity and pressure
fields satisfy the incompressible Navier-Stokes equations
for a fluid with density $\rho_f$ and kinematic
viscosity $\nu$. Turbulence is maintained in a statistically stationary homogeneous
isotropic state via injection of energy at
the lowest wavenumbers\,($1 \leq k_f \leq 2$)\,\cite{samriddhi1}. The
simulations are pseudospectral in space and involve a second-order Adams-Bashforth
scheme for time marching. A spatial resolution of $512^3$ collocation points is
used, with the choice of four different kinematic viscosities
corresponding to Taylor-scale Reynolds numbers, $R_\lambda =
u_{rms}^2\sqrt{15/\epsilon\nu}$, of $47$, $96$, $150$ and $200$\,($u_{rms}$ is
the root-mean-square velocity and $\epsilon=2 \nu \langle
\bm{E}\!\!:\!\!\bm{E}\rangle$ is the averaged dissipation rate). For each
$R_\lambda$, we follow the motion of $100000$ oblate (prolate) spheroids, with
aspect ratios\,($\kappa$) ranging from $0.1$ to $0.01$\,($10$ to $100$); here,
$\kappa = a/b$, $a$ and $b$ being the semi-axis lengths along and orthogonal to
the spheroid symmetry axis $\bm{p}$. The particles are initialized at random
positions with their translational velocities
set equal to fluid values and angular velocities set equal to
those of anisotropic tracers~\cite{Jeffery1922} at their locations. The initial orientations, as characterized by normalized quaternions\,\cite{evans77}, are
uniformly distributed over the unit sphere. The simulations are run for $5-6$
integral-scale eddy turnover times, sufficient to attain a statistical
steady state.

The equations governing the particle dynamics are:
\begin{align}
&\qquad\qquad\qquad\qquad\frac{d{\boldsymbol U}_p}{dt} =  {\bm{g}}\!\!+\!\!\frac{1}{\tau_p X_A}\boldsymbol{M}_t^{-1}\cdot({\boldsymbol u}\!\!-\!\!{\boldsymbol U}_p) \label{eq:translation}, \\&
\frac{d{\boldsymbol \omega}_p}{dt}\!\!+\!\!\bm{I}_p^{-1}\!\cdot\![\bm{\omega}_p \!\wedge\! (\bm{I}_p\!\cdot\!\bm{\omega}_p)] \!\!=\!\! 
K_{sed} \bm{I}_p^{-1}\!\cdot\![(\bm{M}_t\!\cdot\!\hat{\bm{g}})\!\cdot\!\bm{p}(\bm{M}_t\!\cdot\!\hat{\bm{g}})\!\wedge\! \bm{p}] \nonumber \\
                              &+8\pi\mu L^3 \bm{I}_p^{-1}\!\cdot\![{\bm M}_r^{-1}\!\cdot\!(\frac{1}{2}\bm{\Omega}-\bm{\omega}_p)
                              -Y_H (\bm{E}\!\cdot\!\bm{p}) \!\wedge\! \bm{p}],  \label{eq:rotation}
\end{align}
where $\bm{U}_p$ and $\bm{\omega}_p$ are the translational and angular
velocities of the particles, ${\bm g}$ is the gravitational acceleration\,($\hat{\bm g}$ being
the corresponding unit vector), $L$ is the largest particle dimension and $\tau_p$ is the
particle relaxation time(see~\cite{SM}). $\bm{I}_p$ in Eq.\ref{eq:rotation} is the moment of
inertia tensor, while ${\bm M}_t$ and ${\bm M}_r$ denote the Stokesian
translational and rotational mobility tensors for the spheroid, with
$\bm{M}_{t(r)}=X_{A(C)}^{-1}(\kappa)\bm{pp}+Y_{A(C)}^{-1}(\kappa)(\bm{I}-\bm{pp})$,
the principal resistance coefficients\,($X_A-Y_C$) being well known functions
of $\kappa$\,\cite{Kimkarrila}. The large particle-to-fluid density
ratio\,($\rho_p/\rho_f$), relevant to the atmospheric scenario, implies the
neglect of Basset and added mass forces in Eq.\ref{eq:translation}. The particle
Reynolds numbers based on both the Kolmogorov shear rate\,($\dot{\gamma}_\eta =
(\epsilon/\nu)^{\frac{1}{2}}$) and the nominal slip velocity\,($U_s = \tau_p
g$) are assumed small\,($ Re_{\dot{\gamma}_\eta}=\dot{\gamma}_\eta L^2 /\nu,
Re_s =U_{s}L/\nu < 1$), so particles are acted on, at leading order, by the sum
of the gravitational force and quasi-steady Stokes drag proportional to the
slip velocity\,\cite{footnote1}; see \cite{SM}. Since sub-Kolmogorov spheroids experience turbulence as a fluctuating linear flow, the Jeffery
relation~\cite{Kimkarrila,Jeffery1922} is used for the turbulent torque in Eq.\ref{eq:rotation} with the ratio $Y_H/Y_C = (\kappa^2-1)/(\kappa^2+1)$
being the Bretherton constant $B$~\cite{Bretherton1962}.
Eq.\ref{eq:rotation} includes, in addition, the gravity-induced torque
acting to orient an anisotropic particle, sedimenting in a quiescent fluid at
small but finite $Re_s$, broadside-on to
gravity\,\cite{cox1965,khayatcox89,navaneeth2015}; an expression for this
torque was obtained in
\cite{navaneeth2015}. The superposition of the gravity and shear-induced
torques in Eq.\ref{eq:rotation} has been used~\cite{footnote2} earlier to determine
the orientation dynamics of particles sedimenting through simple shear
flow~\cite{subkoch2005, subkoch2006}. The quantity $T\!R =
\frac{K_{sed}}{\mu L^3 \dot{\gamma}_\eta}\sim Fr_\eta^2 f_I(\kappa)$
characterizes the relative magnitudes of these
torques in Eq.\ref{eq:rotation}, where $K_{sed}=Re_s\mu U_s
L^2f_I(\kappa)X_A^2$, with the aspect-ratio dependent function, $f_I(\kappa)$,
having been obtained in \cite{navaneeth2015}, and $Fr_\eta=\tau_p
g/u_{\eta}$ being the Froude number based on the Kolmogorov
velocity scale\ (${\bm u}_\eta = (\nu\epsilon)^{\frac{1}{4}})$. In
Eq.\ref{eq:translation} and Eq.\ref{eq:rotation}, $\bm{u}$, $\bm{\Omega}$ and
$\bm{E}$ denote the undisturbed turbulent velocity, vorticity and rate-of-strain fields
interpolated at the particle positions~\cite{SM}.

\begin{figure*}
\includegraphics[width=1.0\columnwidth]{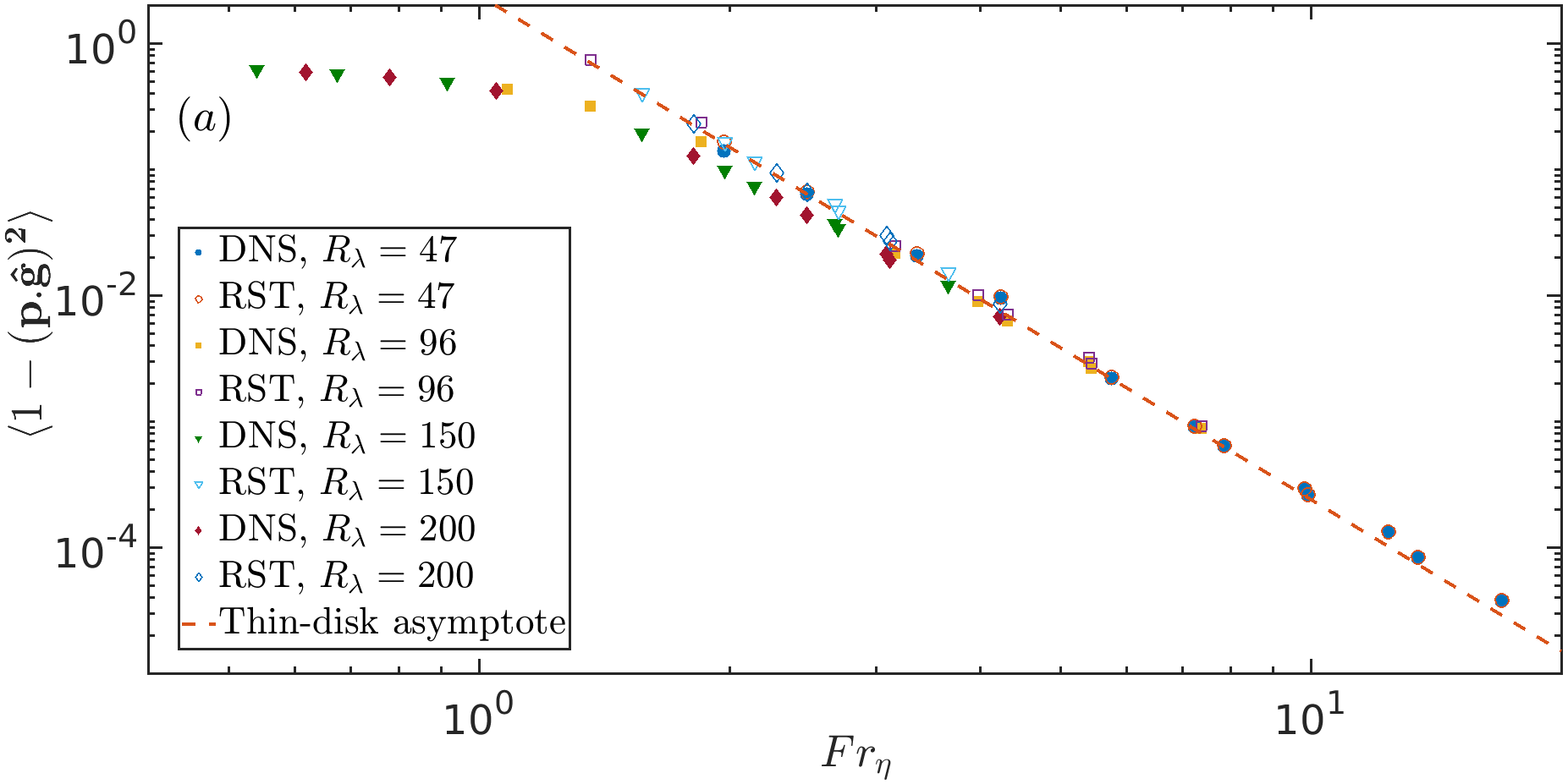}
\includegraphics[width=1.0\columnwidth]{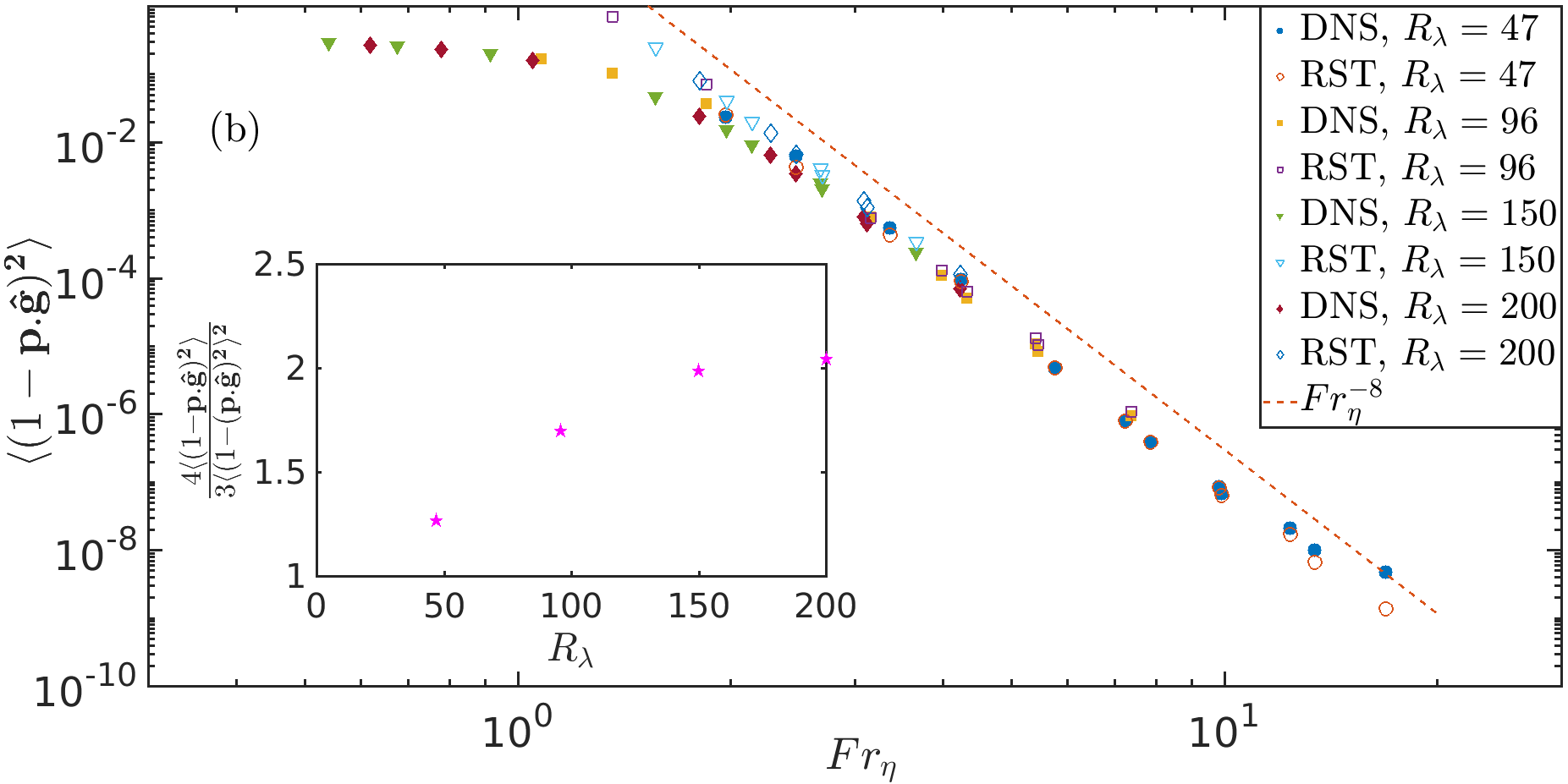}
\caption{ Comparison of the orientation moments (a) $\langle 1- ({\bm p} \cdot \hat{\bm g})^2 \rangle$ and (b) $\langle (1- {\bm p} \cdot \hat{\bm g})^2 \rangle$ 
obtained from DNSs with RST predictions\,(\eqref{eq:secmoment} and \eqref{eq:fourthmoment}) in the rapid-settling limit\,($Fr_\eta \gg 1$). 
The inset in panel (b) is a measure (see text) of the departure from Gaussianity of the orientation distributions.} 
\label{fig:moments}
\end{figure*}

Apart from $R_\lambda$, $\kappa$ and $Fr_\eta$, the dynamics as governed by
Eq.\ref{eq:translation} and Eq.\ref{eq:rotation}, on length scales of the order of the
Kolmogorov scale\,($l_\eta = (\nu^3/\epsilon)^{\frac{1}{4}}$) or smaller, is a function of the Kolmogorov Stokes number\,($St_\eta = \tau_p/\tau_\eta$
with $\tau_\eta=\dot{\gamma}_\eta^{-1}$ the Kolmogorov time scale). Using parameters characteristic of the atmospheric scenario, including ice
crystal sizes and turbulence dissipation rates from \cite{pumir}, the
simulations reported here correspond to $St_\eta \in (0.0037,0.4)$ and $Fr_\eta
\in (0.5,17)$. For a given $R_\lambda$, the dynamics of the
thinnest\,(disk-like) spheroids corresponds to the smallest Stokes and Froude
numbers. The torque
ratio, $T\!R$ ranges from $1-800$ for all ice crystal sizes and
turbulence intensities considered here. Thus, the
gravity-induced torque is expected to be dominant for typical ice
clouds. This is borne out in Fig.~\ref{fig:pdf} which
shows the distribution of orientations $\cos \theta_0 = |\hat{\bm g} \cdot {\bm p}|$ (since ${\bm p}$ and $-{\bm p}$ correspond to the same spheroid
orientation, we take the modulus), obtained from 
our DNSs for (a) $R_\lambda =150$ and (b) $R_\lambda =47$. For each $R_\lambda$, 
we show results for oblate spheroids of different aspect ratios (see legend), both with($Re_s\!=\!0.8$) and without($Re_s\!=\!0$) the gravity-induced torque. The gravity-induced torque causes the distributions to be sharply localized about the broadside-on
orientation ($\cos \theta_0 = 1$), especially for the smaller $R_\lambda$. In contrast,
as emphasized in the insets of Fig.~\ref{fig:pdf}, neglect of this torque leads to distributions peaked at the
longside-on orientation\,($\cos \theta_0 = 0$ for oblate spheroids),
although this maximum is quite shallow, consistent with earlier studies\,\cite{siewert1,siewert2,pumir}. The continuous curves in Fig.~\ref{fig:pdf} are a guide to the eye; the comparison with a Gaussian in Fig.~\ref{fig:pdf}(a) nevertheless conveys the pronouncedly non-Gaussian character of the distributions for $Re_s\!=\!0.8$.

Analytical progress is possible in the rapid-settling limit(henceforth, RST or `Rapid-Settling Theory'), $l_\eta/U_s
\ll \tau_\eta$  or $Fr_\eta \gg 1$, when a particle
settles through a Kolmogorov eddy much faster than the eddy
decorrelates~\cite{anu2019}(see~\cite{SM}). Further, assuming $St_\eta \ll 1$, and
neglecting the angular acceleration in Eq.\ref{eq:rotation}, the rate of change
of spheroid orientation, $\dot{\bm p} = {\bm \omega}_p \wedge \bm{p}$, is given
by:
\begin{align} 
\dot{\bm p} =&-\bm{M}_r\cdot[K_{sed}[(\bm{M}_t\cdot\hat{\bm{g}})\cdot\bm{p}(\bm{M}_t\cdot\hat{\bm{g}})]\wedge \bm{p}] \wedge {\bm p} \nonumber\\
& +\frac{1}{2}\bm{\Omega} \wedge \bm{p}+\frac{Y_H}{Y_c}[\bm{E}\cdot\bm{p}-\bm{E}:\bm{ppp}].  \label{rot:inertialess}
\end{align}

As already seen, the torque ratio $T\!R \sim f_I(\kappa)Fr_\eta^2$ with
$f_I(\kappa) \sim \mathcal{O}(1)$ for oblate spheroids. For large $Fr_\eta$, the weak turbulent shear only leads to
small fluctuations about the broadside-on orientation. For such orientations,
with $\hat{\bm{g}}\!=\!\bm{1}_3$, one has $\bm{p}\cdot\hat{\bm{g}}=p_3 \approx 1$
and $p_{1,2} \ll 1$. Furthermore, the rotation rate of the nearly broadside-on
spheroid, in any plane containing $\hat{\bm g}$, is asymptotically small since
the gravity-induced torque vanishes for the broadside-on orientation. Thus, there is a near-balance
between the $1$ and $2$ components of the turbulent and gravity-induced torques at leading order, the terms
proportional to $\dot{p}_{1,2}$ in (\ref{rot:inertialess}) being $\mathcal{O}(Fr_\eta^{-2})$ smaller. This gives
\begin{align}\bm{p}\cdot(\bm{I}-\hat{\bm{g}}\hat{\bm{g}}) \approx \frac{1}{Fr_\eta^2}\frac{8\pi
Y_A Y_c \tau_\eta}{f_I(\kappa) X_A}(\bm{S}+\frac{Y_H}{Y_c}\bm{E})\cdot\bm{p}
\end{align}
for the projection of the spheroid axis in the plane transverse to gravity; here $\bm{S}=\frac{1}{2}\bm{\epsilon} \cdot \bm{\Omega}$ is the vorticity tensor and $\bm{\epsilon}$ being the Levi-Civita symbol. The components $p_{1,2}$ transverse to gravity are linear functionals of the turbulent velocity gradient tensor.  Turbulent velocity gradients are dominated by the smallest\,(Kolmogorov) scales, and are pronouncedly non-Gaussian\cite{footnote4}; hence the orientation distributions, in the rapid-settling limit, are non-Gaussian(characterized below via the second and fourth moments) despite the localization about the broadside-on orientation.

Since $p_3 = \cos \theta_0 \approx 1- \frac{\theta_0^2}{2}$ for $\theta_0 \ll
1$, $\langle  1-p_3^2 \rangle = \langle p_1^2 + p_2^2 \rangle \approx \langle \theta_0^2 \rangle$ corresponds to the
variance of the orientation distribution about
the broadside-on orientation. With $p_{1,2}$ linear in ${\bm E}$ and ${\bm
S}$, calculating $\langle 1-p_3^2\rangle$ requires the
variance of the turbulent rate of strain and vorticity tensors
over a particle settling trajectory. For $St_\eta \ll 1, Fr_\eta \gg
1$, one expects no preferential sampling and the average along a
settling trajectory, $\langle \cdot \rangle$, above may be replaced by the
usual fluid ensemble average\,\cite{Batchelor1953}. For homogeneous isotropic
turbulence, the ensemble averages are: $\langle E_{ij} E_{kl}\rangle =
\frac{\dot{\gamma}_\eta^2}{20}(\delta_{ik}\delta_{jl} + \delta_{il}\delta_{jk}-
\frac{2}{3}\delta_{ij}\delta_{kl}), $ $\langle S_{ij}S_{kl}\rangle=
\frac{\dot{\gamma}_\eta^2}{12}(\delta_{ik}\delta_{jl} -
\delta_{il}\delta_{jk}), $ and $\langle S_{ij}E_{kl} \rangle =
0$\,\cite{BrunkKoch1997,PopeBook}. Using these~\cite{SM}, one finds:
\begin{align}
\langle 1-p_3^2 \rangle\approx \frac{32\pi^2 Y_A^2 Y_c^2}{f_I^2(\kappa) X_A^2}\left(\frac{1}{3}+\frac{Y_H^2}{5Y_c^2}\right)\frac{1}{Fr_\eta^4}. \label{eq:secmoment} \end{align} 
Fig.~\ref{fig:moments}(a) compares the DNS results for $\langle 1-p_3^2
\rangle$ to Eq.\ref{eq:secmoment} and demonstrates the good agreement for large
$Fr_\eta$, with deviations arising for $Fr_\eta$ of order unity and smaller, in
which case $\langle 1- p_3^2 \rangle$ approaches a plateau.

A more sensitive measure of the orientation distributions is $\langle (1-p_3)^2
\rangle$. For a distribution localized about the broadside-on
orientation, $\langle (1-p_3)^2 \rangle \propto \langle
\theta_0^4 \rangle$, and is therefore a measure of the fourth moment.
Proceeding along lines sketched above, $\langle (1-p_3)^2
\rangle \approx \frac{1}{4}\langle (p_1^2 +p_2^2)^2 \rangle$ with $p_{1,2}$ as given above, and the calculation involves the fourth moment of the turbulent
velocity gradient tensor~\cite{SM}. One obtains:
\begin{eqnarray}
\langle (1\!-\!p_3)^2 \rangle\! \approx \!\! \frac{1}{4}(\!\frac{8\pi Y_A Y_c}{f_I(\kappa) X_A}\!)^4 [M_1\!+\!B^2 M_2\! +\! B^4 M_3]\frac{1}{Fr_\eta^8}, \label{eq:fourthmoment}
\end{eqnarray}
where $M_1=\frac{3G_1}{2}+\frac{32G_2}{15}-\frac{162G_3}{15}$,
$M_2=-3G_1+\frac{8G_2}{15}+\frac{54G_3}{5}$ and $M_3=\frac{3 G_1}{2}$, with
$G_1\!=\!\langle (\tau_\eta\partial{u_1}/\partial{x_1})^4 \rangle$,
$G_2\!=\!\langle (\tau_\eta\partial{u_1}/\partial{x_2})^4 \rangle$ and
$G_3\!=\!\langle \tau_\eta^4(\partial{u_1}/\partial{x_1})^2
(\partial{u_1}/\partial{x_2})^2\rangle$ being the independent (non-dimensional)
scalar components involving the fourth moment of the velocity gradient.
Unlike the second moment, the pre-factor multiplying $Fr_\eta^{-8}$
is both a function of $\kappa$ and $R_\lambda$, the latter dependence arising
from dissipation-range intermittency referred to above. Fig.
~\ref{fig:moments}(b) compares Eq.\ref{eq:fourthmoment} with
DNS results, the pattern of agreement being similar to that of the second
moment above\,\cite{footnote3}. Since $\langle 1-p_3^2 \rangle = \langle
\theta_o^2 \rangle$ and $\langle (1-p_3)^2 \rangle = \frac{1}{4}\langle
\theta_o^4 \rangle$ for large $Fr_\eta$, the ratio $\frac{4\langle (1-p_3)^2
\rangle}{3\langle 1-p_3^2 \rangle^2}$, which is independent of $Fr_\eta$,
characterizes the departure from
Gaussianity. This ratio, which is unity for a Gaussian, is plotted as
an inset in Fig.~\ref{fig:moments}(b) for $\kappa \rightarrow 0$\,(a flat
disk); it is well above unity and increases with increasing $R_\lambda$. One therefore expects orientation distributions in the atmospheric case, with $R_\lambda$'s one to two orders of magnitude higher than those in our simulations\,~\cite{shaw2003}, to have similar variances but be significantly more intermittent.

\begin{figure}
\includegraphics[width=\columnwidth]{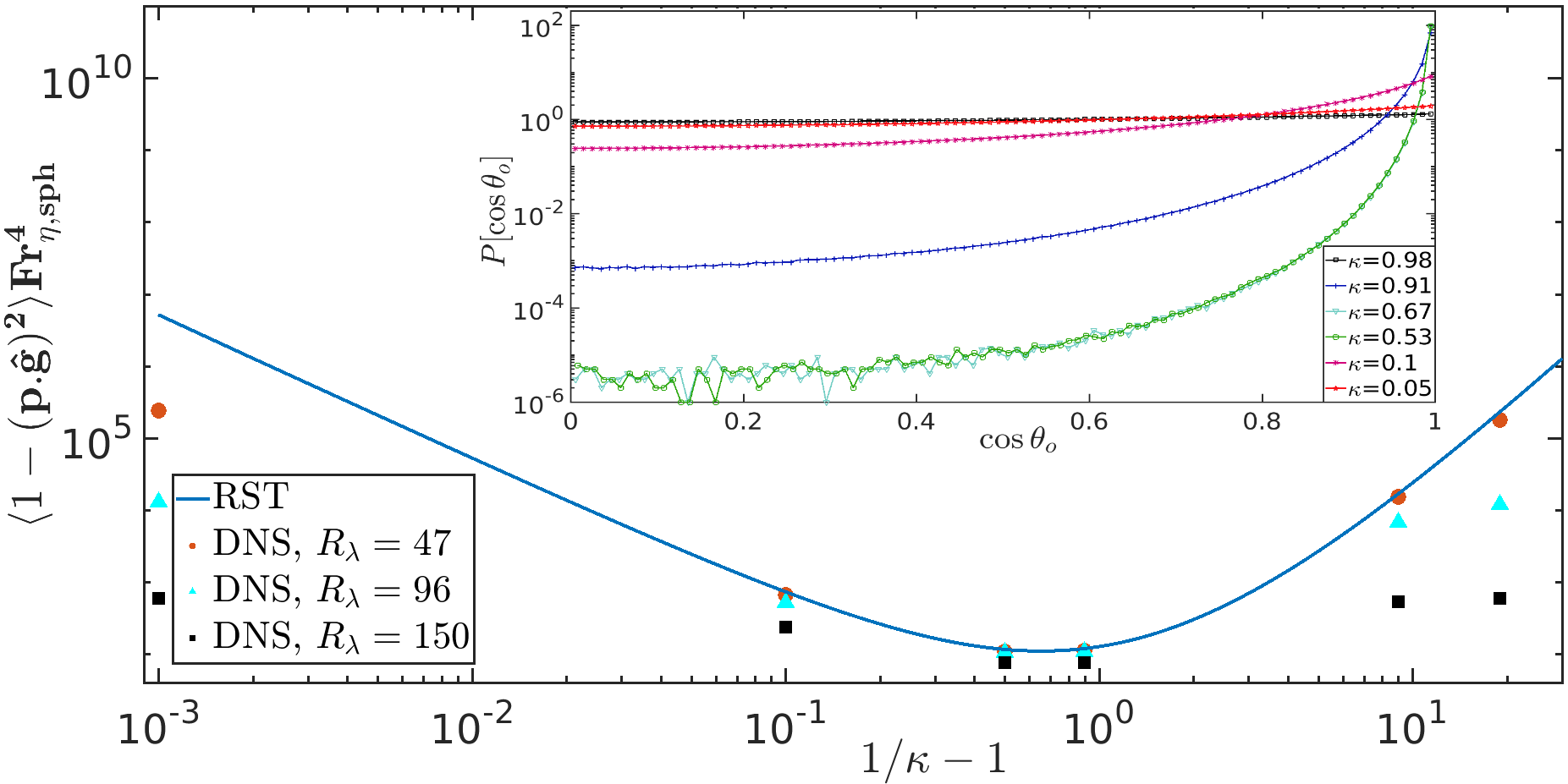}
\caption {Orientation distributions for $R_\lambda=96$ for various aspect ratios, all other parameters staying fixed\,(see \cite{SM}). The inset highlights the non-monotonic behavior of the second moment, $\langle 1-(\bm{p\cdot\hat{g}})^2 \rangle$, scaled with $Fr_{\eta,sph}^4$, when plotted as a function of $\kappa$ for $R_\lambda = 47, 96$ and $150$. } 
\label{fig:momentAspectratio}
\end{figure}

 In the inset of Fig.~\ref{fig:momentAspectratio}, we plot orientation distributions as a function of the spheroid aspect ratio, other physical parameters being fixed\,\cite{SM}. Interestingly, the localization about the broadside-on orientation first increases as $\kappa$ increases from zero\,(a flat disk), attains a maximum, before decreasing again as $\kappa$ approaches unity. The non-monotonicity arises because the gravity-induced torque is small for both flat disks\,(due to the vanishingly small mass of such shapes) and near-spheres\,(since the torque scales with the square of the small eccentricity). The second moment from the RST framework, Eq.\ref{eq:secmoment}\!, can be rewritten to isolate the $\kappa$-dependence through a change of variable $Fr_\eta=Fr_{\eta,sph}*\frac{\kappa}{X_A}$,  
where $Fr_{\eta,sph} = \frac{2\rho_p L^2 g}{9 \mu u_\eta}$. The resulting $\kappa-$dependence is consistent with the above non-monotonicity; although, within the RST framework, $\langle 1-p_3^2 \rangle \sim \mathcal{O}(\kappa^{-4})$ for $\kappa \rightarrow 0$ and $\langle 1-p_3^2 \rangle \sim \mathcal{O}(\kappa-1)^{-2}$ for $\kappa \rightarrow 1$. Since $\langle 1-p_3^2 \rangle \leq 1$, the divergences above betray a breakdown of the assumption of a localized distribution in the analysis. As shown in Fig.~\ref{fig:momentAspectratio}, the second moments from our DNS agree with Eq.\ref{eq:secmoment} for intermediate values of $\kappa$\,(maximum localization of $\cos\theta_0$), but plateau in the aforementioned asymptotic limits\,(corresponding to a uniform distribution of $\cos\theta_0$). Overall, the disagreement with theory, expectedly, grows with increasing $R_\lambda$.

\begin{figure}
\includegraphics[width=1.0\columnwidth]{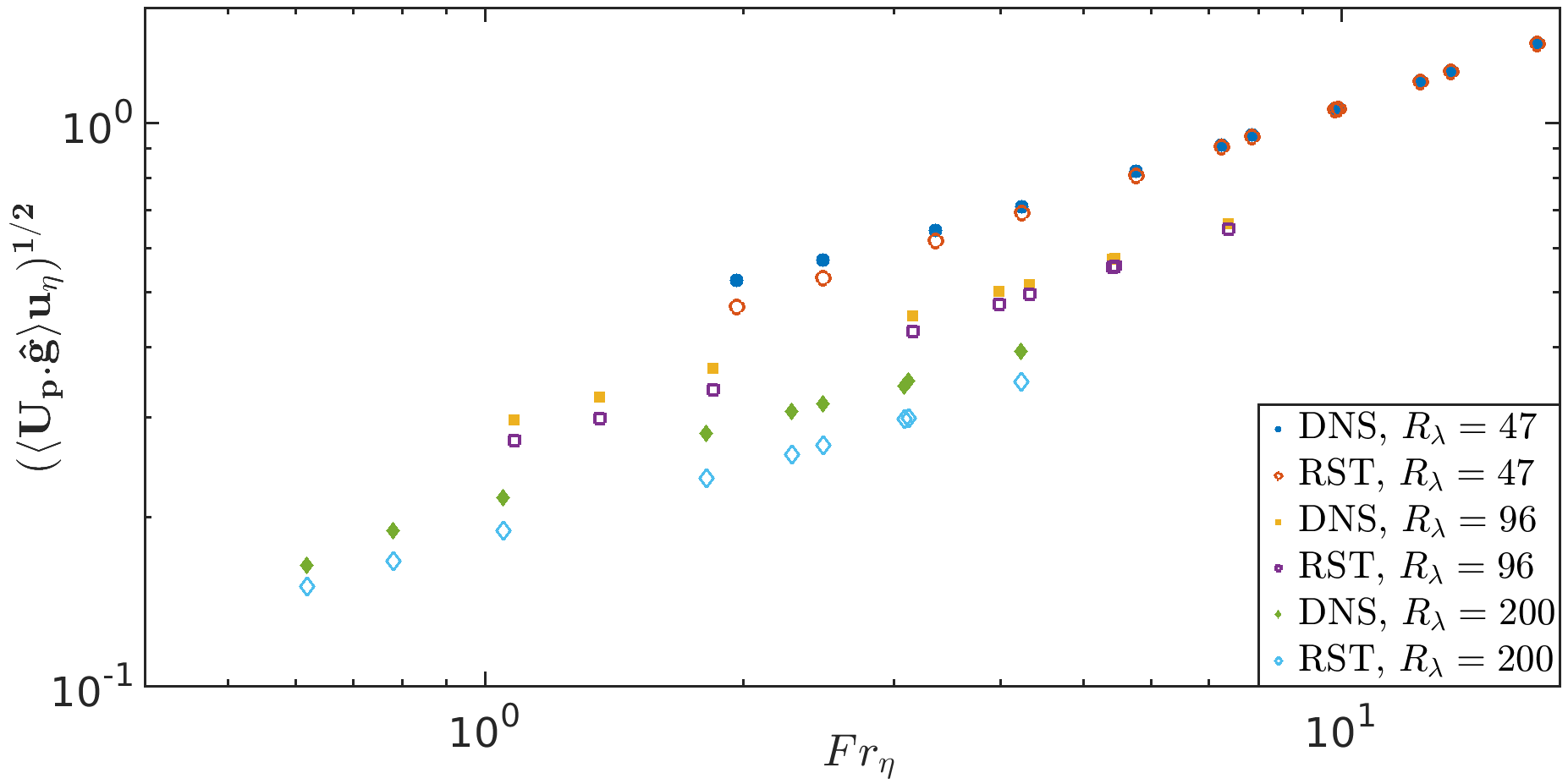}
\caption{Comparison between time-averaged and orientation-averaged settling speeds for $R_\lambda \!=\! 47$, $96$ and $200$.}
\label{fig:vsettling}
\end{figure}

With increase in the turbulence intensity, $Fr_\eta$ decreases while
$St_\eta$ increases to values of order unity. As already seen in Fig.~\ref{fig:moments}, DNS results depart from
RST predictions in this limit. A suspension of spherical particles in a turbulent flow is no longer spatially homogeneous when $Fr_\eta, St_\eta \sim \mathcal{O}(1)$\,\cite{BecRay2014,eatonReview,Collins2016a,Collins2016b}. Preferential sampling of regions of low vorticity by inertial particles, together with a
sweeping effect in presence of gravity, leads to
enhanced settling velocities\,\cite{maxey_87,maxey_93,collins2014}. Fig.~\ref{fig:vsettling}
shows this to be true for the suspensions of spheroids considered here.
For large $Fr_\eta$, the time-averaged settling speeds(which scale linearly with $Fr_\eta$ on account of being proportional to the acceleration due to gravity) from the DNS agree with
the orientational averages for $R_\lambda = 47$ and $96$\,(the $Fr_\eta$ required for this agreement increases with increasing $St_\eta$). For finite $Fr_\eta$ and
$St_\eta$, the time averages consistently exceed the orientation averaged
estimates due to the preferential sweeping effect\,\cite{SM}.

In this letter, we have characterized the orientation distributions and settling speeds of spheroids in homogeneous isotropic turbulence. Orientation distributions are localized about the broadside-on\,(to gravity) orientation, but are pronouncedly non-Gaussian for parameters typical of the atmospheric scenario. This is in contrast to recent studies which neglect the gravity-induced torque, and predict distributions peaked at the longside-on orientation\,\cite{pumirmehlig2017,siewert1,siewert2,pumir}. The non-Gaussian distributions found here are also in contrast to earlier analyses reliant on a Gaussian ansatz\,\cite{Klett1994,mehligNJP2019}. 
 While the broadside-on peak has been captured in \cite{mehligNJP2019}, the simplistic Gaussian ansatz used for the velocity field, and the resulting Gaussian nature of the orientation fluctuations, is incorrect. Furthermore, \cite{mehligNJP2019} lacks any discussion on the spatial organization of the particles, and its effect on particle settling speeds. In contrast, we show that the particle concentration field remains homogeneous for $St_\eta \ll 1$;  for $St_\eta \sim \mathcal{O}(1)$, preferential sweeping effects lead to a spatially inhomogeneous concentration and enhanced settling speeds\,(Fig.~\ref{fig:vsettling},\cite{SM}). Results for prolate spheroids (not shown) are similar to those discussed above. It would be of interest, in future, to characterize pair-level statistics for anisotropic particles in position-orientation space, as a step towards analyzing ice-water and ice-ice collision efficiencies; the latter thought of as crucial to explaining observed ice-crystal concentrations in mixed-phase clouds and relatively rapid snow-flake formation in ice clouds\,\cite{Khain1,Khain2,Khain3}.

\begin{acknowledgments}
SSR acknowledges the financial support of the DAE, Govt. of India, under project no. 12-R\&D-TFR-5.10-1100 and DST (India) project ECR/2015/000361. The simulations were performed on the ICTS clusters {\it Mowgli} and {\it Mario}, as well as
the work stations from the project ECR/2015/000361: {\it Goopy} and
{\it Bagha}.
\end{acknowledgments}


\clearpage
\onecolumngrid

\begin{center}
  \textbf{\large Supplemental material}\\[.2cm]
\end{center}

\setcounter{equation}{0}
\setcounter{figure}{0}
\setcounter{table}{0}
\setcounter{page}{1}
\renewcommand{\theequation}{S\arabic{equation}}
\renewcommand{\thefigure}{S\arabic{figure}}
\renewcommand{\bibnumfmt}[1]{[S#1]}
\renewcommand{\citenumfont}[1]{S#1}
\section{The spheroid equations of motion}
The equations governing particle translation and rotation are given by:
\begin{align}
&\qquad\qquad\qquad\qquad\frac{d{\boldsymbol U}_p}{dt} =  {\bm{g}}\!\!+\!\!\frac{1}{\tau_p X_A}\boldsymbol{M}_t^{-1}\cdot({\boldsymbol u}\!-\!{\boldsymbol U}_p) \label{eq:Stranslation}, \\&
\frac{d{\boldsymbol \omega}_p}{dt}\!\!+\!\!\bm{I}_p^{-1}\!\cdot\![\bm{\omega}_p \!\wedge\! (\bm{I}_p\!\cdot\!\bm{\omega}_p)] \!\!=\!\! 
K_{sed} \bm{I}_p^{-1}\!\cdot\![(\bm{M}_t\!\cdot\!\hat{\bm{g}})\!\cdot\!\bm{p}(\bm{M}_t\!\cdot\!\hat{\bm{g}})\!\wedge\! \bm{p}] \nonumber \\
                              &+8\pi\mu L^3 \bm{I}_p^{-1}\!\cdot\![{\bm M}_r^{-1}\!\cdot\!(\frac{1}{2}\bm{\Omega}-\bm{\omega}_p)
                              -Y_H (\bm{E}\!\cdot\!\bm{p}) \!\wedge\! \bm{p}].  \label{eq:Srotation}
\end{align}

The Stokesian translational\,($\bm{M}_{t}$) and rotational\,($\bm{M}_{r}$) mobility tensors appearing in (\ref{eq:Stranslation}) and (\ref{eq:Srotation}) characterize the viscous force and torque acting on the spheroid. For a spheroid, these are of the form $\bm{M}_{t(r)}=X_{A(C)}^{-1}(\kappa)\bm{pp}+Y_{A(C)}^{-1}(\kappa)(\bm{I}-\bm{pp})$, with the scalar aspect-ratio dependent resistance functions given as follows \cite{Kim}:
\begin{itemize}
\item Oblate:
\begin{subequations} \label{eq:oblateresist}
\begin{align}
 X_A&=\dfrac{4(1-\kappa^2)^{3/2}}{3(\kappa(1-\kappa^2)^{1/2}+(1-2\kappa^2)\cos^{-1}(\kappa))}, \\
 Y_A&=-\dfrac{8(1-\kappa^2)^{3/2}}{3(\kappa(1-\kappa^2)^{1/2}+(-3+2\kappa^2)\cos^{-1}(\kappa))}, \\
 X_c&=-\dfrac{2(1-\kappa^2)^{3/2}}{3(\kappa(1-\kappa^2)^{1/2}-\cos^{-1}(\kappa))}, \\
 Y_c&=-\dfrac{2(1-\kappa^2)^{1/2}(\kappa^4-1)}{3(\kappa(1-\kappa^2)^{1/2}+(1-2\kappa^2)\cos^{-1}(\kappa))}. \\
\end{align}
\end{subequations}
\item Prolate:
\begin{subequations} \label{eq:prolateresist}
\begin{align}
 X_A&=\dfrac{8(\kappa^2-1)^{3/2}}{3\kappa((2 \kappa ^2-1) \log (\frac{\kappa+(\kappa ^2-1)^{1/2}}{\kappa-(\kappa ^2-1)^{1/2}})-2 \kappa(\kappa^2-1)^{1/2})}, \\
 Y_A&=\dfrac{16(\kappa^2-1)^{3/2}}{3\kappa((2\kappa ^2-3) \log (-1+2\kappa(\kappa+(\kappa^2-1)^{1/2}))+2 \kappa(\kappa^2-1)^{1/2})}, \\
 X_c&=\dfrac{4(\kappa^2-1)^{3/2}}{3\kappa^3(2 \kappa(\kappa^2-1)^{1/2}-\log(\frac{\kappa+(\kappa ^2-1)^{1/2}}{\kappa-(\kappa^2-1)^{1/2}}))}, \\
 Y_c&=\dfrac{4(\kappa^2-1)^{3/2}(\kappa^2+1)}{3\kappa^3((2 \kappa^2-1) \log (\frac{\kappa+(\kappa^2-1)^{1/2}}{\kappa-(\kappa^2-1)^{1/2}})-2\kappa(\kappa^2-1)^{1/2})}. \\
\end{align}
\end{subequations}
\end{itemize}
Since sub-Kolmogorov spheroids experience turbulence as a fluctuating linear flow, the Jeffery relation~\cite{Kim,jeffery} has been used for the turbulent shear torque in (\ref{eq:Srotation}) with the ratio $Y_H/Y_C = (\kappa^2-1)/(\kappa^2+1)$ being the Bretherton constant $B$~\cite{Bretherton_1962}.
The particle relaxation time, $\tau_p$, appearing in equation (\ref{eq:Stranslation}) is defined as $\tau_p=\dfrac{2 \rho_p b^2 \kappa}{9 \mu X_A}$ for oblate($\kappa<1$) and $\tau_p=\dfrac{2 \rho_p a^2}{9 \mu \kappa^2 X_A}$ for prolate($\kappa>1$) spheroids. The moment of inertia tensor in equation (\ref{eq:Srotation}) is defined by:
\begin{itemize}
\item Oblate:
\begin{eqnarray}
 \bm{I}_p &=& \dfrac{4\pi\rho_p b^5 \kappa}{15}[2\,\bm{pp}+ (1+\kappa^2)(\bm{I}-\bm{pp})],
\end{eqnarray}
\item Prolate:
\begin{eqnarray}
 \bm{I}_p &=& \dfrac{4\pi\rho_p a^5}{15\kappa^4} [2\,\bm{pp}+ (1+\kappa^2)(\bm{I}-\bm{pp})].
\end{eqnarray}
\end{itemize}

There is an additional and important torque contribution in (\ref{eq:Srotation}) due to gravity, arising from the effects of fluid inertia associated with a spheroid settling in an otherwise quiescent fluid. The dominant contribution to this torque is due to inertial forces acting in a region around the spheroid of order its own size, and the torque is therefore proportional to the sedimentation Reynolds number\,($Re_s$) for $Re_s \ll 1$. In other words, the gravity-induced inertial torque emerges as a regular perturbation about the Stokesian limit. Therefore, to $O(Re_s)$, the functional dependence of the torque on $\hat{\bm g}$ and ${\bm p}$, viz, the form $(\bm{M}_t\!\cdot\!\hat{\bm{g}})\!\cdot\!\bm{p}(\bm{M}_t\!\cdot\!\hat{\bm{g}})\!\wedge\! \bm{p}$ in (\ref{eq:Srotation}) may be readily inferred using symmetry arguments. The additional aspect ratio dependence, contained in the coefficient $K_{sed}$ in (\ref{eq:Srotation}), requires a detailed analysis. This calculation has been done in \cite{navaneeth_2015}, using a generalized reciprocal theorem formulation, and one obtains $K_{sed} =Re_s\mu U_s
L^2 f_I(\kappa)X_A^2$, with the aspect-ratio dependent inertial function, $f_I(\kappa)$ specified below:
\begin{itemize}
	\item Oblate:
	\begin{align}
	f_I(\kappa) &= \pi (1-\kappa ^2)^2 [4757 \kappa ^8-9039 \kappa ^6+2075 \kappa ^4+4359 \kappa ^2-2152 \nonumber \\
	& + 210 (20 \kappa ^6-13 \kappa ^4-12 \kappa ^2+7) (1-\kappa ^2)^{1/2} \kappa  \sin ^{-1}((1-\kappa ^2)^{1/2})  \nonumber \\     
	& - 105 (24 \kappa ^6-55 \kappa ^4+50 \kappa ^2-19) \sin ^{-1}((1-\kappa ^2)^{1/2})^2]/D,
	\end{align}
	where
	$D=315 (\kappa (\kappa ^2-1)-(1-\kappa ^2)^{1/2} (2 \kappa ^2-3) \sin ^{-1}((1-\kappa ^2)^{1/2}))\\
	(\kappa (\kappa ^2-1)+(2 \kappa ^2-1) (1-\kappa ^2)^{1/2} \sin ^{-1}((1-\kappa ^2)^{1/2}))^2$
	\item Prolate:
	\begin{align}
	f_I(\kappa) &=\pi (\kappa ^2-1) [-2207 (\kappa ^2-1)^{1/2} \kappa^2 +2152 (\kappa ^2-1)^{1/2}\nonumber \\ 
	&+105 (24 \kappa ^2-31) (\kappa ^2-1)^{1/2} \kappa^2  \tanh ^{-1}(\frac{(\kappa ^2-1)^{1/2}}{\kappa})^2\nonumber \\
	&+1995 (\kappa ^2-1)^{1/2} \tanh ^{-1}(\frac{(\kappa ^2-1)^{1/2}}{\kappa})^2\nonumber \\
	&+4757(\kappa ^2-1)^{1/2} \kappa ^6-4282 (\kappa ^2-1)^{1/2} \kappa ^4 \nonumber \\
	&-210\kappa (20 \kappa ^6-13 \kappa ^4-12 \kappa ^2+7) \tanh^{-1}(\frac{(\kappa ^2-1)^{1/2}}{\kappa})]/E,
	\end{align}
\end{itemize}
where $E=315 \kappa ^3 (\kappa  (\kappa ^2-1)^{1/2}+(1-2 \kappa ^2) \tanh ^{-1}(\frac{(\kappa ^2-1)^{1/2}}{\kappa}))^2\\ (\kappa  (\kappa ^2-1)^{1/2}+(2 \kappa ^2-3) \tanh ^{-1}(\frac{(\kappa ^2-1)^{1/2}}{\kappa}))$.\\

On account of inertia being a regular perturbation, use of the generalized reciprocal theorem shows that the inertial correction to the viscous torque in (\ref{eq:Srotation}), in the limit $Re_s, Re_{\dot{\gamma}_\eta} \ll 1$\,($Re_{\dot{\gamma}_\eta}$ being the Reynolds number based on the Kolmogorov shear rate), may be constructed as a linear superposition\,\cite{subkoch_2005,subkoch_2006} of a shear-induced contribution in the absence of gravity ($\sim F_I(\kappa)Re_{\dot{\gamma}_\eta}\mu L^3 \dot{\gamma_{\eta}}$, see \cite {navaneeth_2016, navaneeth_2017}) and the gravity-induced contribution above that neglects any ambient shear. The ratio of the turbulent shear-induced inertial to the gravity-induced torques then turns out to be $\mathcal{O}(Fr_\eta^{-2})(L/l_\eta)^2 \ll 1$. Thus, the shear-induced inertial torque may be neglected for the sub-Kolmogorov spheroids examined here, and only the gravity-induced contribution is therefore included in (\ref{eq:Srotation}). 

It is worth noting that, in contrast to the torque problem, the inertial correction to the Stokes drag in (\ref{eq:Stranslation}) is not a linear superposition of the gravity and shear contributions since inertial effects enter as a singular perturbation in this case. This is evident from the \cite{saffman_1965} and \cite{mclaughlin_1991} derivations for the inertial lift on sphere in a simple shear flow. Even in the limit $Re_s, Re_{\dot{\gamma}_\eta} \ll 1$, the scaling and the direction of the inertial force is crucially dependent on the ratio of the two screening length\,($\nu/U_s$ for sedimentation and $(\nu/\dot{\gamma}_\eta)^{\frac{1}{2}}$ for shear). Importantly, however, both of these inertial screening lengths are much larger than the size of the (sub-Kolmogorov) spheroid, and the inertial corrections, although non-trivial, are nevertheless small in comparison to the Stokes drag in (\ref{eq:Stranslation}). Non-linear corrections to the drag become important for particles comparable to the Kolmogorov scale; these corrections are not known in closed form, however, and one then needs fully resolved simulations~\cite{schneider_2019}.

Based on the above system of equations, we perform direct numerical simulations of non-interacting spheroids sedimenting through homogeneous and isotropic turbulence with a mass loading assumed small enough for carrier-fluid turbulence to remain unaffected; that is to say, a one-way coupled framework. For mass loadings of order unity, one requires a two-way coupled framework~\cite{horwitz_2016, bala_2017}, in which case the fluid velocity needs to be accurately determined at the particle positions in order to estimate the particle forces (which then act to modify the turbulence); this in turn involves rather subtle issues with regard to interpolation schemes~\cite{horwitz_2016, bala_2017}.

\section{The Rapid-Settling Theory\,(RST)}
\subsection{Formulation}
We turn our attention to equation (\ref{eq:Srotation}), which governs the rotational dynamics of the particles. Our objective is to calculate $\bm{\omega}_p$ in the limit where the particles settle rapidly through a Kolmogorov eddy in a time much smaller than the eddy decorrelation time; that is, $l_\eta/U_s \ll \tau_\eta \Rightarrow U_s/u_\eta$ or $Fr_\eta \gg 1$. We also assume that the Stokes number based on the Kolmogorov timescale,  $St_\eta \ll 1$. The angular acceleration of the particles, in equation (\ref{eq:Srotation}), can then be neglected since it is $\mathcal{O}(St_\eta/Fr_\eta^{2})$ smaller than the gravity-induced torque and $\mathcal{O}(St_\eta)$ smaller than the turbulent shear-induced torque. The rotation rate of a spheroid is given by:
\begin{eqnarray}
 \dot{\bm{p}}=\bm{\omega}_p \wedge \bm{p}, \label{eq:pdot}
\end{eqnarray}
The contributions to $\bm{\omega}_p$, and hence $\dot{\bm{p}}$, are $\dot{\bm{p}}^{sed}$ from the gravity-induced torque and $\dot{\bm{p}}^{jeff}$ from the turbulent shear-induced torque.

To calculate $\dot{\bm{p}}^{sed}$, we use the Stokesian relation between the gravity-induced torque and the hydrodynamic torque acting on a spheroid rotating in a quiescent fluid, which yields:
\begin{align}
 K_{sed} (M_{(t)kl} \hat{g}_l p_k)(\epsilon_{imn}M_{(t)mj} \hat{g}_j p_n)=-8\pi\mu L^3 M^{-1}_{(r)ij} \omega_{pj}^{sed}. 
\end{align}
Using $\hat{\bm{g}}=-\bm{1}_3$ in the above expression, one obtains:
\begin{eqnarray}
 \omega_{pj}^{sed} &=& -\dfrac{f_I(\kappa)\tau_p^2 g^2 X_A }{8\pi\nu Y_A Y_c} \epsilon_{j3n} p_3 p_n.
\end{eqnarray}
Substituting in (\ref{eq:pdot}),
\begin{align}
 \dot{p}_i^{sed}&=-\dfrac{f_I(\kappa)\tau_p^2 g^2 X_A}{8\pi\nu Y_A Y_c}p_3(p_3 p_i-\delta_{i3}). \label{eq:gravity}
\end{align}

In the sub-Kolmogorov range, particles see the turbulence as a fluctuating linear flow. In the absence of inertial effects, the particles rotate with an angular velocity given by the Jeffery relation at leading order. This gives:
\begin{align}
 \dot{p}_{i}^{jeff}&=(S_{ji}+B E_{ji})p_j-B E_{jk}p_jp_kp_i, \label{eq:jeffery}
\end{align}
where $B$ is the Bretherton's constant, as defined earlier.
Adding the gravity-induced~(\ref{eq:gravity}) and the Jeffery~(\ref{eq:jeffery}) contributions, one obtains:
\begin{align}
 \dot{p}_i&=(S_{ji}+BE_{ji})p_j-BE_{jk}p_jp_kp_i-\dfrac{f_I(\kappa)\tau_p^2 g^2 X_A }{8\pi\nu Y_A Y_c} p_3(p_i p_3-\delta_{i3}). \label{rot:inertialess}
\end{align}
\newline
For large $Fr_\eta$, the dominant gravity-induced torque implies that, for both oblate and prolate spheroids, the weak turbulent shear only leads to small fluctuations about the broadside-on orientation\,(for oblate spheroids, the broadside equilibrium corresponds to $p_3 = 1, p_{1,2}=0$; for prolate spheroids $p_3 = 0$). The rotation rate of a nearly broadside-on oblate spheroid($p_{1,2}\ll p_3 \approx 1$), in any plane containing $\hat{\bm g}$, is asymptotically small, owing to the vicinity to the aforementioned gravity-induced equilibrium. Thus, in the $1$ and $2$ components of (\ref{rot:inertialess}) for oblate spheroids, there is a near-balance between the turbulent and gravity-induced torques at leading order; the terms proportional to $\dot{p}_{1,2}$ are $\mathcal{O}(p_{1,2}\dot{\gamma}_\eta)$ with the Jeffery contribution being $\mathcal{O}(p_{3}\dot{\gamma}_\eta)$. Since $p_3 \approx 1$, and $p_{1,2}$ turn out to be $\mathcal{O}(Fr_\eta^{-2})$\,(see (\ref{variance_p1}) and (\ref{variance_p2}) below), the unsteady terms may be neglected with an error of $\mathcal{O}(Fr_\eta^{-2})$. The term $\bm{E}:\bm{ppp}$ in (\ref{rot:inertialess}) is also ignored since it involves quadratic combinations of $p_1$ and $p_2$, all of which are asymptotically small in the rapid settling limit, as seen above. With these approximations, one obtains:
\begin{align}
p_1 &\approx \dfrac{1}{Fr_\eta^2}\dfrac{8\pi Y_A Y_c \tau_\eta}{f_I(\kappa) X_A}(S_{31} + \frac{Y_H}{Y_c}E_{31})\label{eq:p1},\\
p_2 &\approx \dfrac{1}{Fr_\eta^2}\dfrac{8\pi Y_A Y_c \tau_\eta}{f_I(\kappa) X_A}(S_{32} + \frac{Y_H}{Y_c}E_{32}) \label{eq:p2},
\end{align}
for an oblate spheroid. Similarly, for a nearly broadside-on prolate spheroid\,($p_3 \ll p_{1,2}\approx \mathcal{O}(1)$), the turbulent and gravity-induced torques nearly balance, at leading order, in the $3^{rd}$ component of (\ref{rot:inertialess}), and one obtains:
\begin{align}
p_3 &\approx \dfrac{1}{Fr_\eta^2}\dfrac{8\pi Y_A Y_c \tau_\eta}{f_I(\kappa) X_A}(S_{13}+S_{23} + \frac{Y_H}{Y_c}(E_{13}+E_{23})) \label{eq:p3}.
\end{align}  

The moments of interest (with regard to characterizing the orientation distribution) are of the general form $\langle(1-p_3^2)^n\rangle$ for an oblate spheroid. Note that the relations (\ref{eq:p1}) and (\ref{eq:p2}) above, which express $p_{1,2}$ as linear functionals of the turbulent velocity gradient, imply that the orientational moment of order $n$ requires the $2n^{th}$ moment of the turbulent velocity gradient. In the rapid settling limit, one expects no preferential sampling, and hence, the averages involved in the aforementioned moments, which are along a settling particle trajectory, may be replaced by the usual fluid ensemble averages(\cite{batchelor1953theory}). The first two moments $\langle 1-p_3^2 \rangle$ and $\langle (1-p_3^2)^2\rangle$, for an oblate spheroid, will be evaluated below. The rapid settling limit corresponds to a small-$\theta_0$ approximation ($\theta_0$ is the angle between the spheroid axis $\bm{p}$ and $\hat{\bm{g}}$), in which case $\langle 1-p_3^2 \rangle \propto \langle\theta_0^2\rangle$ and $\langle (1-p_3^2)^2\rangle \propto \langle\theta_0^4\rangle$. Thus, evaluating these will also allow one to characterize the departure of the distribution from Gaussianity.

\subsection{The second moments of the turbulent velocity gradient and spheroid orientation}
The second moment $\langle 1-p_3^2 \rangle=\langle p_1^2+p_2^2\rangle$, and we therefore begin by squaring equations (\ref{eq:p1}), (\ref{eq:p2}) and then ensemble averaging. The latter eliminates terms proportional $\langle\bm{SE}\rangle$ from symmetry arguments pertaining to homogeneous isotropic turbulence. This gives:
\begin{subequations} \label{eq:psq}
\begin{align} 
  \langle p_1^2 \rangle \approx \dfrac{32\pi^2 Y_A^2 Y_c^2}{f_I^2(\kappa) X_A^2}(\langle S_{ji}S_{km}\rangle+\dfrac{Y_H^2}{Y_c^2}\langle E_{ji}E_{km}\rangle)p_jp_k\delta_{i1}\delta_{m1}\dfrac{1}{Fr_\eta^4}, \label{variance_p1} \\
  \langle p_2^2 \rangle \approx \dfrac{32\pi^2 Y_A^2 Y_c^2}{f_I^2(\kappa) X_A^2}(\langle S_{ji}S_{km}\rangle+\dfrac{Y_H^2}{Y_c^2}\langle E_{ji}E_{km}\rangle)p_j p_k\delta_{i2}\delta_{m2}\dfrac{1}{Fr_\eta^4}. \label{variance_p2}
\end{align}
\end{subequations}
For an oblate spheroid, the largest terms in the double contraction in equation \ref{eq:psq} are those that involve $p_3\,(\approx 1)$, and (\ref{variance_p1}) and (\ref{variance_p2}) reduce to. 
\begin{subequations} \label{eq:psq2}
\begin{align} 
  \langle p_1^2 \rangle \approx \dfrac{32\pi^2 Y_A^2 Y_c^2}{f_I^2(\kappa) X_A^2}(\langle S_{31}S_{31}\rangle+\dfrac{Y_H^2}{Y_c^2}\langle E_{31}E_{31}\rangle)\dfrac{1}{Fr_\eta^4},\\
  \langle p_2^2 \rangle \approx \dfrac{32\pi^2 Y_A^2 Y_c^2}{f_I^2(\kappa) X_A^2}(\langle S_{32}S_{32}\rangle+\dfrac{Y_H^2}{Y_c^2}\langle E_{32}E_{32}\rangle)\dfrac{1}{Fr_\eta^4}. 
\end{align}
\end{subequations}
Evaluating (\ref{eq:psq2}) requires the variance of the turbulent velocity gradient tensor which is given by\,\cite{Pope_Book}:
\begin{align}
 \langle \Gamma_{ji} \Gamma_{lk}\rangle&=\frac{2{\dot{\gamma}^2}_\eta}{15}(\delta_{ik}\delta_{jl}-\frac{1}{4}\delta_{ij}\delta_{kl}-\frac{1}{4}\delta_{il}\delta_{jk}).  \label{eq:GG}
\end{align}
The expressions for the variance of the rate-of-strain tensor, $\bm{E}=\frac{\bm{\Gamma}+\bm{\Gamma}^\dagger}{2}$ and the vorticity tensor, $\bm{S}=\frac{\bm{\Gamma}-\bm{\Gamma}^\dagger}{2}$ can be obtained from \ref{eq:GG}. For example,
\begin{align}
 \langle S_{31}S_{31}\rangle&=\dfrac{1}{4}[\langle \Gamma_{31}\Gamma_{31}\rangle-\langle \Gamma_{31}\Gamma_{13}\rangle-\langle \Gamma_{13}\Gamma_{31}\rangle+\langle \Gamma_{13}\Gamma_{13}\rangle]\nonumber \\
 &=\frac{{\dot{\gamma}^2}_\eta}{12},\nonumber \\
 \langle E_{31}E_{31}\rangle&=\dfrac{1}{4}[\langle \Gamma_{31}\Gamma_{31}\rangle+\langle \Gamma_{31}\Gamma_{13}\rangle+\langle \Gamma_{13}\Gamma_{31}\rangle+\langle \Gamma_{13}\Gamma_{13}\rangle]\nonumber \\
 &=\frac{{\dot{\gamma}^2}_\eta}{20}
\end{align}


Since all directions in the plane perpendicular to gravity are equivalent, one has $\langle S_{31}S_{31}\rangle = \langle S_{32}S_{32}\rangle$ and $\langle E_{31}E_{31}\rangle = \langle E_{32}E_{32}\rangle$. The explicit expressions for $\langle\bm{EE}\rangle$ and $\langle\bm{SS}\rangle$ have, in fact, already been given in \cite{brunk_koch}, but we nevertheless calculate them using the full velocity-gradient tensor $\bm{\Gamma}$, since this serves as a prelude to the fourth moment derivation in Section~\ref{sec:fourthmoment}).

Adding equations \ref{eq:psq2}a and \ref{eq:psq2}b and using equation (\ref{eq:GG}) to calculate the ensemble averages, $\langle 1-p_3^2 \rangle$ for an oblate spheroid turns out to be:
\begin{align}
  \langle 1-p_3^2 \rangle \approx \dfrac{32\pi^2 Y_A^2 Y_c^2}{f_I^2(\kappa) X_A^2}(\dfrac{1}{3}+\dfrac{Y_H^2}{Y_c^2} \dfrac{1}{5})\dfrac{1}{Fr_\eta^4}. \label{eq:secondmoment}
\end{align}
Proceeding along similar lines, the second moment for a nearly broadside-on prolate spheroid, $\langle p_3^2 \rangle$, is shown to be half of (\ref{eq:secondmoment}).

\subsection{Fourth-moment of the turbulent velocity gradient}

In this section, we evaluate the orientational moment $\langle (1-p_3)^2 \rangle \approx \frac{1}{4}\langle (1-p_3^2)^2 \rangle$ which, as indicated above, is proportional to the fourth moment of the orientation distribution when localized about the broadside-on equilibrium. The calculation involves first obtaining the fourth moment of the turbulent velocity gradient, and this involves a rather elaborate effort. We use a graphical approach that allows substantial simplification of the algebra involved. In order to illustrate the approach, we derive expressions for both the third and fourth moments of the velocity gradient tensor, defined below:
\begin{itemize}
\item Third moment:
\begin{eqnarray}
 \Sigma_{ijkpqr}=\langle \Gamma_{pi}\Gamma_{qj}\Gamma_{rk}\rangle 
               =\langle \frac{\partial{u_i}}{\partial{x_p}} \frac{\partial{u_j}}{\partial{x_q}} \frac{\partial{u_k}}{\partial{x_r}} \rangle. \label{eq:thirdmoment}
\end{eqnarray}
\item Fourth moment: 
\begin{eqnarray}
 \Pi_{ijkmsrqp}=\langle \Gamma_{si}\Gamma_{rj}\Gamma_{qk}\Gamma_{pm}\rangle 
               =\langle \dfrac{\partial{u_i}}{\partial{x_s}} \dfrac{\partial{u_j}}{\partial{x_r}} \dfrac{\partial{u_k}}{\partial{x_q}} \dfrac{\partial{u_m}}{\partial{x_p}} \rangle. \label{eq:fourthmoment}
\end{eqnarray}
\end{itemize}
Note that the result for the third moment of the velocity gradient tensor, $\Sigma_{ijkpqr}$, is already known\,(see page 206 in \cite{Pope_Book}), but is nevertheless rederived here for purposes of clarity. 

The sixth-order tensor $\Sigma_{ijkpqr}$ and eighth-order tensor $\Pi_{ijkmsrqp}$ must evidently be isotropic, and therefore expressible in terms of tensor products of the Kronecker delta. The usual derivation involves writing down all possible permutations of the Kronecker deltas, although this becomes especially tedious for the fourth and higher moments. To circumvent the algebraic effort involved, we use an alternate method where the aforementioned permutations are represented as `graphs', with terms corresponding to the same graphs being grouped together. Starting off with $\Sigma_{ijkpqr}$, we note that the graph of every term in $\Sigma_{ijkpqr}$ is composed of three lines, each of these connecting a pair of indices in (\ref{eq:thirdmoment}). Thus, each line corresponds to a Kronecker delta tensor in the final permutation sum. There are three types of lines:
 \begin{itemize}
  
  \item A vertical line \begin{tikzpicture}
        \draw[thick] (0.25,0.2) -- (0.25,-0.2);
        \end{tikzpicture} connects indices belonging to the same partial derivative in $\Sigma_{ijkpqr}$. For example, since the indices `i' and `p' occur in the same partial derivative $\dfrac{\partial{u_i}}{\partial{x_p}}$, $\delta_{ip}$ is represented as 
        \begin{tikzpicture}
        \draw[thick] (0.25,0.2) -- (0.25,-0.2);
        \end{tikzpicture}.
  \item A slant line\,\begin{tikzpicture}
        \draw[thick] (0.2,0.2) -- (-0.2,-0.2);
        \end{tikzpicture}
 connects two indices in $\Sigma_{ijkpqr}$ belonging to a $\bm{u}$ and an $\bm{x}$ that correspond to different partial derivatives. For example, the index `i' in $\Sigma_{ijkpqr}$ occurs in $\dfrac{\partial{u_i}}{\partial{x_p}}$ while `q' occurs in $\dfrac{\partial{u_j}}{\partial{x_q}}$. So, $\delta_{iq}$ is written as
        \begin{tikzpicture}
        \draw[thick] (0.2,0.2) -- (-0.2,-0.2);
        \end{tikzpicture}.
  \item A horizontal line\, \begin{tikzpicture}
        \draw[thick] (-0.3,4) -- (0.3,4);
        \end{tikzpicture} connects indices in $\Sigma_{ijkpqr}$ belonging either to two $\bm{u}$'s or two $\bm{x}$'s; the indices involved obviously correspond to different partial derivatives. For example, the index `i' in $\Sigma_{ijkpqr}$ occurs on $\dfrac{\partial{u_i}}{\partial{x_p}}$ while `m' occurs on $\dfrac{\partial{u_k}}{\partial{x_r}}$, and $\delta_{ik}$ is thus written as
        \begin{tikzpicture}
        \draw[thick] (-0.3,4) -- (0.3,4);
        \end{tikzpicture}
        
 \end{itemize}
The above definitions lead to five distinct graphs for $\Sigma_{ijkpqr}$. Thus, the fifteen terms in the original permutation sum may be divided into five groups with elements in a given group having the same graph. Each of the groups is multiplied by a scalar constant, so the effort reduces to determining five rather than fifteen different constants. The different graphs, along with a representative element corresponding to each one, may be stated as follows: 
	 $\delta_{ip}\delta_{jq}\delta_{kr}$: 	    
	    \begin{tikzpicture}
	    \draw[thick] (0.25,0.2) -- (0.25,-0.2);
	    \draw[thick] (0.12,.2) -- (0.12,-0.2);
	    \draw[thick] (-0.01,0.2) -- (-0.01,-0.2);
	    \end{tikzpicture},	    
	 $\delta_{ip}\delta_{jk}\delta_{qr}$: 	    
	    \begin{tikzpicture}
	    \draw[thick] (0.25,0.2) -- (0.25,-0.2);
	    \draw[thick] (0.35,.1) -- (0.75,.1);
	    \draw[thick] (0.35,-.1) -- (0.75,-.1);
	    \end{tikzpicture},	    
	 $\delta_{ip}\delta_{jr}\delta_{qk}$: 	    
	    \begin{tikzpicture}
	    \draw[thick] (0.25,0.2) -- (0.25,-0.2);
	    \draw[thick] (0.35,-.2) -- (0.6,.2);
	    \draw[thick] (0.5,-.2) -- (0.75,.2);
	    \end{tikzpicture},	    
       	 $\delta_{iq}\delta_{pk}\delta_{jr}$: 	    
	    \begin{tikzpicture}
	    \draw[thick] (0.35,-.2) -- (0.6,.2);
	    \draw[thick] (0.5,-.2) -- (0.75,.2);
	    \draw[thick] (0.65,-.2) -- (0.9,0.2);
	    \end{tikzpicture},	    
	 $\delta_{ij}\delta_{pk}\delta_{qr}$: 	    
	    \begin{tikzpicture}
	    \draw[thick] (0.2,-.1) -- (0.6,-.1);
	    \draw[thick] (0.2,.1) -- (0.6,.1);
	    \draw[thick] (0.6,-.2) -- (0.85,.2);
	    \end{tikzpicture}\,(note that the ordering of the lines in a graph does not matter. For instance, $\delta_{ip}\delta_{jk}\delta_{qr}$ can be represented by either    
	    \begin{tikzpicture}
	    \draw[thick] (0.25,0.2) -- (0.25,-0.2);
	    \draw[thick] (0.35,.1) -- (0.75,.1);
	    \draw[thick] (0.35,-.1) -- (0.75,-.1);
	    \end{tikzpicture}
	    or
	    \begin{tikzpicture}
            \draw[thick] (-0.15,-.1) -- (0.25,-.1);
	    \draw[thick] (0.35,0.2) -- (0.35,-0.2);
  	    \draw[thick] (-0.15,.1) -- (0.25,.1);
	    \end{tikzpicture}\,).
	    
In light of the above, $\Sigma_{ijkpqr}$ can be written as:
\begin{align}
\Sigma_{ijkpqr} &= c_1(\,
            \begin{tikzpicture}
	    \draw[thick] (0.25,0.2) -- (0.25,-0.2);
	    \draw[thick] (0.12,.2) -- (0.12,-0.2);
	    \draw[thick] (-0.01,0.2) -- (-0.01,-0.2);
	    \end{tikzpicture}\,)   
            +c_2(\,    
	    \begin{tikzpicture}
	    \draw[thick] (0.25,0.2) -- (0.25,-0.2);
	    \draw[thick] (0.35,.1) -- (0.75,.1);
	    \draw[thick] (0.35,-.1) -- (0.75,-.1);
	    \end{tikzpicture}\,)
	    +c_3(\,
	    \begin{tikzpicture}
	    \draw[thick] (0.25,0.2) -- (0.25,-0.2);
	    \draw[thick] (0.35,-.2) -- (0.6,.2);
	    \draw[thick] (0.5,-.2) -- (0.75,.2);
	    \end{tikzpicture}\,)
	    +c_4(\,
	    \begin{tikzpicture}
	    \draw[thick] (0.35,-.2) -- (0.6,.2);
	    \draw[thick] (0.5,-.2) -- (0.75,.2);
	    \draw[thick] (0.65,-.2) -- (0.9,0.2);
	    \end{tikzpicture}\,)
	    +c_5(\,
	    \begin{tikzpicture}
	    \draw[thick] (0.2,-.1) -- (0.6,-.1);
	    \draw[thick] (0.2,.1) -- (0.6,.1);
	    \draw[thick] (0.6,-.2) -- (0.85,.2);
	    \end{tikzpicture}\,)
\end{align}
In actual tensorial notation, this becomes:
\begin{align}
 \Sigma_{ijkpqr} &=c_1 K^{(1)}_{ipjqkr} + c_2 K^{(2)}_{ipjkqr} + c_3 K^{(3)}_{ipjrqk} + c_4 K^{(4)}_{iqpkjr} + c_5 K^{(5)}_{ijpkqr}, \label{eq:thirdmomentdef}
\end{align}
  where the $K^{(n)}$'s are defined as:
 \begin{subequations} \label{eq:thirdmomentdef2} 
 \begin{align}
  K^{(1)}_{ipjqkr} &= \delta_{ip}\delta_{jq}\delta_{kr}, \\
  K^{(2)}_{ipjkqr} &= \delta_{ip}\delta_{jk}\delta_{qr} + \delta_{jq}\delta_{ik}\delta_{pr} + \delta_{kr}\delta_{ij}\delta_{pq},\\
  K^{(3)}_{ipjrqk} &= \delta_{ip}\delta_{jr}\delta_{qk} + \delta_{jq}\delta_{ir}\delta_{pk} + \delta_{kr}\delta_{iq}\delta_{pj},\\
  K^{(4)}_{iqpkjr} &= \delta_{iq}\delta_{pk}\delta_{jr} + \delta_{ir}\delta_{pj}\delta_{qk},\\
  K^{(5)}_{ijpkqr} &= \delta_{ij}\delta_{pk}\delta_{qr} + \delta_{ij}\delta_{qk}\delta_{pr} + \delta_{ik}\delta_{pj}\delta_{qr} + \delta_{ik}\delta_{rj}\delta_{pq} + \delta_{jk}\delta_{qi}\delta_{pr} + \delta_{jk}\delta_{ri}\delta_{pq}.
\end{align}
\end{subequations}
 
$\Sigma_{ijkpqr}$ is invariant to certain indicial permutations, and the grouping on the right hand side of  (\ref{eq:thirdmomentdef}) is consistent with these invariances. Now, from continuity, one has $\Sigma_{ijkiqr}=0$, which leads to:
\begin{subequations} \label{eq:incomprthird}
\begin{align} 
 3c_1+2c_2+2c_3&=0, \\
 3c_2+4c_5&=0, \\
 3c_3+2c_4+2c_5&=0.
\end{align}
\end{subequations}

Next, we make use of the homogeneity condition~\cite{Pope_Book}: 
\begin{align}
  \dfrac{\partial{ }}{\partial{x_i}}\langle u_k\frac{\partial{u_i}}{\partial{x_j}}\frac{\partial{u_j}}{\partial{x_k}}\rangle&=\langle \dfrac{\partial{u_k}}{\partial{x_i}}\dfrac{\partial{u_i}}{\partial{x_j}}\dfrac{\partial{u_j}}{\partial{x_k}}\rangle=0,
\end{align}
which leads to:
\begin{align}  
 c_1+3c_2+9c_3+10c_4+12c_5&=0 \label{eq:homogenthird}
\end{align}

The four relations between the $c_i$'s above imply that there is a single scalar that characterizes $\Sigma_{ijkpqr}$. Based on these four relations, one finds $c_2=-\dfrac{4}{3}c_1,\,c_3=-\dfrac{1}{6}c_1,\, c_4=-\dfrac{3}{4}c_1$ and $c_5=c_1$, and the constant $c_1$ may in turn be expressed in terms of a particular scalar third moment of the turbulent velocity gradient. A convenient (and standard) choice is the skewness based on the longitudinal velocity gradient $\langle(\tau_\eta\frac{\partial{u_1}}{\partial{x_1}})^3\rangle$, which is known to be negative (on account of vortex stretching) for homogeneous isotropic turbulence. Thus, the final expression for the third moment of the turbulent velocity gradient reads as:
\begin{align}
 \Sigma_{ijkpqr} &=& \langle(\tau_\eta\frac{\partial{u_1}}{\partial{x_1}})^3\rangle[K^{(1)}_{ipjqkr} -\frac{4}{3} K^{(2)}_{ipjkqr} -\frac{1}{6} K^{(3)}_{ipjrqk} - \frac{3}{4} K^{(4)}_{iqpkjr} + K^{(5)}_{ijpkqr}], \label{eq:thirdmomentfinal}
\end{align}
The average, $\langle(\tau_\eta\frac{\partial{u_1}}{\partial{x_1}})^3\rangle$, which is dependent on $R_\lambda$, may now computed from DNS.

The procedure outlined above is now followed for calculating the fourth moment $\Pi_{ijkmsrqp}$. The 105 possible permutations of the Kronecker deltas may be organized into eight groups with all elements in a group again having the same graph. Thus, the calculation reduces to determining only eight constants, and moreover, not all of these are independent (as already seen above, there are constraints imposed from continuity and homogeneity). Thus, $\Pi_{ijkmsrqp}$ may be written as:
\begin{align}
\Pi_{ijkmsrqp} &= c_1(\,
            \begin{tikzpicture}
	    \draw[thick] (0.25,0.2) -- (0.25,-0.2);
	    \draw[thick] (0.12,.2) -- (0.12,-0.2);
	    \draw[thick] (-0.01,0.2) -- (-0.01,-0.2);
	    \draw[thick] (-0.14,0.2) -- (-0.14,-0.2);
	    \end{tikzpicture}\,)   
            +c_2(\,    
	    \begin{tikzpicture}
	    \draw[thick] (0.12,.2) -- (0.12,-0.2);
	    \draw[thick] (0.25,0.2) -- (0.25,-0.2);
	    \draw[thick] (0.35,-.1) -- (0.75,-.1);
	    \draw[thick] (0.35,.1) -- (0.75,.1);
	    \end{tikzpicture}\,)
	    +c_3(\,
	    \begin{tikzpicture}
	    \draw[thick] (0.12,.2) -- (0.12,-0.2);
	    \draw[thick] (0.25,0.2) -- (0.25,-0.2);
	    \draw[thick] (0.35,-.2) -- (0.6,.2);
	    \draw[thick] (0.5,-.2) -- (0.75,.2);
	    \end{tikzpicture}\,)
	    +c_4(\,
	    \begin{tikzpicture}
	    \draw[thick] (0.25,0.2) -- (0.25,-0.2);
	    \draw[thick] (0.35,-.2) -- (0.6,.2);
	    \draw[thick] (0.5,-.2) -- (0.75,.2);
	    \draw[thick] (0.65,-.2) -- (0.9,0.2);
	    \end{tikzpicture}\,)
	    +c_5(\,
	    \begin{tikzpicture}
	    \draw[thick] (0.12,0.2) -- (0.12,-0.2);
	    \draw[thick] (0.2,-.1) -- (0.6,-.1);
	    \draw[thick] (0.2,.1) -- (0.6,.1);
	    \draw[thick] (0.6,-.2) -- (0.85,.2);
	    \end{tikzpicture}\,)
	    \nonumber \\&+c_6(\,
	    \begin{tikzpicture}
	    \draw[thick] (0.2,-0.2) -- (0.45,0.2);
	    \draw[thick] (0.35,-.2) -- (0.6,.2);
	    \draw[thick] (0.5,-.2) -- (0.75,.2);
	    \draw[thick] (0.65,-.2) -- (0.9,.2);
	    \end{tikzpicture}\,)
	    +c_7(\,
	    \begin{tikzpicture}
	    \draw[thick] (0.05,-.1) -- (0.45,-.1);
	    \draw[thick] (0.05,.1) -- (0.45,.1);
	    \draw[thick] (0.5,-0.2) -- (0.75,0.2);
	    \draw[thick] (0.65,-.2) -- (0.9,.2);
	    \end{tikzpicture}\,)
	    +c_8(\,
	    \begin{tikzpicture}
	    \draw[thick] (0.05,-.1) -- (0.45,-.1);
	    \draw[thick] (0.05,.1) -- (0.45,.1);
	    \draw[thick] (0.5,-0.1) -- (0.9,-0.1);
	    \draw[thick] (0.5,.1) -- (0.9,.1);
	    \end{tikzpicture}\,),
\end{align}
or, in terms of tensorial notation:
\begin{align}
 \Pi_{ijkmsrqp}&=c_1 K^{(1)}_{isjrkqmp} + c_2 K^{(2)}_{isjrkmpq} + c_3 K^{(3)}_{isjrkpmq} + c_4 K^{(4)}_{isjpkrmq} + c_5 K^{(5)}_{isjqkmpr} \nonumber \\& + c_6 K^{(6)}_{irjskpmq} + c_7 K^{(7)}_{irjskmpq}+c_8 K^{(8)}_{imjkpsqr}, \label{eq:fourthmomentdef}
\end{align}
  where the $K^{(n)}$'s are defined as:
 \begin{subequations} \label{eq:fourthmomentdef2} 
 \begin{align}
  K^{(1)}_{isjrkqmp} &= \delta_{is}\delta_{jr}\delta_{kq}\delta_{mp}, \\
  K^{(2)}_{isjrkmpq} &= \delta_{is}\delta_{jr}\delta_{km}\delta_{pq} + \delta_{is}\delta_{jm}\delta_{kq}\delta_{pr} + \delta_{is}\delta_{jk}\delta_{mp}\delta_{qr} + \delta_{im}\delta_{jr}\delta_{kq}\delta_{ps} + \delta_{ik}\delta_{jr}\delta_{mp}\delta_{qs}\nonumber\\& + \delta_{ij}\delta_{kq}\delta_{mp}\delta_{rs},\\
  K^{(3)}_{isjrkpmq} &= \delta_{is}\delta_{jr}\delta_{kp}\delta_{mq} + \delta_{is}\delta_{jq}\delta_{kr}\delta_{mp}+\delta_{is}\delta_{jp}\delta_{kq}\delta_{mr} +\delta_{ir}\delta_{js}\delta_{kq}\delta_{mp} + \delta_{iq}\delta_{jr}\delta_{ks}\delta_{mp}\nonumber\\&+ \delta_{ip}\delta_{jr}\delta_{kq}\delta_{ms},\\
  K^{(4)}_{isjpkrmq} &= \delta_{is}\delta_{jp}\delta_{kr}\delta_{mq} +\delta_{is}\delta_{jq}\delta_{kp}\delta_{mr}+ \delta_{ir}\delta_{jq}\delta_{ks}\delta_{mp} +\delta_{ir}\delta_{jp}\delta_{kq}\delta_{ms} + \delta_{iq}\delta_{js}\delta_{kr}\delta_{mp} \nonumber\\& +\delta_{iq}\delta_{jr}\delta_{kp}\delta_{ms} +\delta_{ip}\delta_{js}\delta_{kq}\delta_{mr}+\delta_{ip}\delta_{jr}\delta_{ks}\delta_{mq},\\
  K^{(5)}_{isjqkmpr} &= \delta_{is}\delta_{jq}\delta_{km}\delta_{pr} + \delta_{is}\delta_{jp}\delta_{km}\delta_{qr}+\delta_{is}\delta_{jm}\delta_{kr}\delta_{pq} + \delta_{is}\delta_{jm}\delta_{kp}\delta_{qr} + \delta_{is}\delta_{jk}\delta_{mr}\delta_{pq}\nonumber\\& + \delta_{is}\delta_{jk}\delta_{mq}\delta_{pr}+ \delta_{ir}\delta_{jm}\delta_{kq}\delta_{ps} + \delta_{ir}\delta_{jk}\delta_{mp}\delta_{qs} + \delta_{iq}\delta_{jr}\delta_{km}\delta_{ps} + \delta_{iq}\delta_{jk}\delta_{mp}\delta_{rs}\nonumber\\& + \delta_{ip}\delta_{jr}\delta_{km}\delta_{qs} + \delta_{ip}\delta_{jm}\delta_{kq}\delta_{rs} +\delta_{im}\delta_{js}\delta_{kq}\delta_{pr} + \delta_{im}\delta_{jr}\delta_{ks}\delta_{pq} + \delta_{im}\delta_{jr}\delta_{kp}\delta_{qs}\nonumber\\& + \delta_{im}\delta_{jp}\delta_{kq}\delta_{rs} + \delta_{ik}\delta_{js}\delta_{mp}\delta_{qr} + \delta_{ik}\delta_{jr}\delta_{ms}\delta_{pq} + \delta_{ik}\delta_{jr}\delta_{mq}\delta_{ps} + \delta_{ik}\delta_{jq}\delta_{mp}\delta_{rs}\nonumber\\& +\delta_{ij}\delta_{ks}\delta_{mp}\delta_{qr} +  \delta_{ij}\delta_{kr}\delta_{mp}\delta_{qs} + \delta_{ij}\delta_{kq}\delta_{ms}\delta_{pr} +\delta_{ij}\delta_{kq}\delta_{mr}\delta_{ps}, \\  
  K^{(6)}_{irjskpmq} &= \delta_{ir}\delta_{js}\delta_{kp}\delta_{mq} +\delta_{ir}\delta_{jq}\delta_{kp}\delta_{ms}+\delta_{ir}\delta_{jp}\delta_{ks}\delta_{mq} + \delta_{iq}\delta_{js}\delta_{kp}\delta_{mr} + \delta_{iq}\delta_{jp}\delta_{ks}\delta_{mr}\nonumber\\& + \delta_{iq}\delta_{jp}\delta_{kr}\delta_{ms} + \delta_{ip}\delta_{js}\delta_{kr}\delta_{mq} + \delta_{ip}\delta_{jq}\delta_{ks}\delta_{mr} + \delta_{ip}\delta_{jq}\delta_{kr}\delta_{ms},\\ 
  K^{(7)}_{irjskmpq} &= \delta_{ir}\delta_{js}\delta_{km}\delta_{pq} + \delta_{ir}\delta_{jq}\delta_{km}\delta_{ps}+\delta_{ir}\delta_{jp}\delta_{km}\delta_{qs} +\delta_{ir}\delta_{jm}\delta_{ks}\delta_{pq} + \delta_{ir}\delta_{jm}\delta_{kp}\delta_{qs}\nonumber\\& + \delta_{ir}\delta_{jk}\delta_{ms}\delta_{pq} +  \delta_{ir}\delta_{jk}\delta_{mq}\delta_{ps} + \delta_{iq}\delta_{js}\delta_{km}\delta_{pr} + \delta_{iq}\delta_{jp}\delta_{km}\delta_{rs} + \delta_{iq}\delta_{jm}\delta_{ks}\delta_{pr}\nonumber\\& + \delta_{iq}\delta_{jm}\delta_{kr}\delta_{ps} + \delta_{iq}\delta_{jm}\delta_{kp}\delta_{rs} + \delta_{iq}\delta_{jk}\delta_{ms}\delta_{pr} + \delta_{iq}\delta_{jk}\delta_{mr}\delta_{ps} + \delta_{ip}\delta_{js}\delta_{km}\delta_{rq}\nonumber\\&+  \delta_{ip}\delta_{jq}\delta_{km}\delta_{rs}+\delta_{ip}\delta_{jm}\delta_{ks}\delta_{qr}+ \delta_{ip}\delta_{jm}\delta_{kr}\delta_{qs} + 
  \delta_{ip}\delta_{jk}\delta_{ms}\delta_{qr} + \delta_{ip}\delta_{jk}\delta_{mr}\delta_{qs}\nonumber\\& + \delta_{ip}\delta_{jk}\delta_{mq}\delta_{rs} + \delta_{im}\delta_{js}\delta_{kr}\delta_{pq} + \delta_{im}\delta_{js}\delta_{kp}\delta_{qr} + \delta_{im}\delta_{jq}\delta_{ks}\delta_{pr} + \delta_{im}\delta_{jq}\delta_{kr}\delta_{ps}\nonumber\\& + \delta_{im}\delta_{jq}\delta_{kp}\delta_{rs} +\delta_{im}\delta_{jp}\delta_{ks}\delta_{qr}+\delta_{im}\delta_{jp}\delta_{kr}\delta_{qs} + \delta_{ik}\delta_{js}\delta_{mr}\delta_{pq} + \delta_{ik}\delta_{js}\delta_{mq}\delta_{pr}\nonumber \\& + 
  \delta_{ik}\delta_{jq}\delta_{ms}\delta_{pr} + \delta_{ik}\delta_{jq}\delta_{mr}\delta_{ps} + \delta_{ik}\delta_{jp}\delta_{ms}\delta_{qr} + 
  \delta_{ik}\delta_{jp}\delta_{mr}\delta_{qs} + \delta_{ik}\delta_{jp}\delta_{mq}\delta_{rs}\nonumber \\& + \delta_{ij}\delta_{ks}\delta_{mr}\delta_{pq} +\delta_{ij}\delta_{ks}\delta_{mq}\delta_{pr} + \delta_{ij}\delta_{kr}\delta_{ms}\delta_{pq} + \delta_{ij}\delta_{kr}\delta_{mq}\delta_{ps} +\delta_{ij}\delta_{kp}\delta_{ms}\delta_{qr}\nonumber\\& + \delta_{ij}\delta_{kp}\delta_{mr}\delta_{qs} + \delta_{ij}\delta_{kp}\delta_{mq}\delta_{rs}, \\  
  K^{(8)}_{imjkpsqr} &= \delta_{im}\delta_{jk}\delta_{ps}\delta_{qr} +\delta_{im}\delta_{jk}\delta_{pr}\delta_{qs}+\delta_{im}\delta_{jk}\delta_{pq}\delta_{rs} + \delta_{ik}\delta_{jm}\delta_{ps}\delta_{qr} + \delta_{ik}\delta_{jm}\delta_{pr}\delta_{qs}\nonumber\\& + \delta_{ik}\delta_{jm}\delta_{pq}\delta_{rs} +\delta_{ij}\delta_{km}\delta_{ps}\delta_{qr} + \delta_{ij}\delta_{km}\delta_{pr}\delta_{qs} + \delta_{ij}\delta_{km}\delta_{pq}\delta_{rs} . 
 \end{align}
\end{subequations}

As was the case for the third moment, the expression (\ref{eq:fourthmomentdef}) is consistent with indicial symmetries of $\Pi_{ijkmsrqp}$. Continuity implies $\Pi_{ijkmirqp}=0$, which leads to:
\begin{subequations} \label{eq:incompr}
\begin{align} 
 c_1+c_2+c_3&=0, \\
 3c_2+4c_5+c_7+c_8&=0, \\
 3c_3+2c_4+2c_5+c_6+c_7&=0, \\
 c_4+c_6+c_7&=0, \\
 3c_5+5c_7+c_8&=0.
\end{align}
\end{subequations}
Next, the homogeneity condition, $\dfrac{\partial{ }}{\partial{x_i}}\langle u_m\dfrac{\partial{u_i}}{\partial{x_j}}\dfrac{\partial{u_j}}{\partial{x_k}}\dfrac{\partial{u_k}}{\partial{x_m}}\rangle=0$, after some manipulation, leads to the following relation:
\begin{align}
\langle\dfrac{\partial{u_m}}{\partial{x_i}}\dfrac{\partial{u_i}}{\partial{x_j}}\dfrac{\partial{u_j}}{\partial{x_k}}\dfrac{\partial{u_k}}{\partial{x_m}}\rangle&=\dfrac{1}{2}\langle (\dfrac{\partial{u_i}}{\partial{x_j}}\dfrac{\partial{u_j}}{\partial{x_i}})^2\rangle.
\end{align}
Rather unexpectedly, the above relation turns out to be an identity based on the relations already known above from continuity. In other words, the homogeneity constraint does not lead to new relations between the scalar constants $c_n$'s. This implies that, of the original eight, three constants are independent, and evaluating them requires three independent (non-dimensional) scalar combinations of four velocity gradients. These are conveniently chosen as:
\begin{subequations} \label{eq:dnsdata}
\begin{align} 
    G_1&=\langle(\tau_\eta\dfrac{\partial{u_1}}{\partial{x_1}})^4\rangle= c_1+6c_2+6c_3+8c_4+24c_5+9c_6+42c_7+9c_8,  \\
    G_2&=\langle(\tau_\eta\dfrac{\partial{u_1}}{\partial{x_2}})^4\rangle= 9c_8,  \\
    G_3&=\langle\tau_\eta^4(\dfrac{\partial{u_1}}{\partial{x_1}})^2(\dfrac{\partial{u_1}}{\partial{x_2}})^2\rangle= c_2+4c_5+7c_7+3c_8 . 
\end{align}
\end{subequations}
Solving (\ref{eq:incompr}) and~(\ref{eq:dnsdata}) leads to: 
\begin{subequations} \label{eq:fourthmomentsol}
\begin{align} 
 c_1&=-\dfrac{1}{8}G_1+\dfrac{8}{45}G_2-\dfrac{9}{10}G_3,\\
 c_2&=-\dfrac{7}{45}G_2+\dfrac{17}{20}G_3,\\
 c_3&=\dfrac{1}{8}G_1-\dfrac{1}{45}G_2+\dfrac{1}{20}G_3,\\
 c_4&=-\dfrac{3}{8}G_1-\dfrac{7}{45}G_2+\dfrac{27}{20}G_3,\\
 c_5&=\dfrac{1}{9}G_2-\dfrac{3}{4}G_3,\\
 c_6&=\dfrac{3}{8}G_1+\dfrac{11}{45}G_2-\dfrac{9}{5}G_3,\\
 c_7&=-\dfrac{4}{45}G_2+\dfrac{9}{20}G_3,\\
 c_8&=\dfrac{1}{9}G_2.
\end{align}
\end{subequations}
The final expression for the fourth moment of the velocity gradient therefore reads as:
\begin{align}
&\Pi_{ijkmsrqp}=\frac{G_1}{8}[-K^{(1)}_{isjrkqmp}+K^{(3)}_{isjrkpmq}-3 K^{(4)}_{isjpkrmq}+3 K^{(6)}_{irjskpmq}]\nonumber\\
&+\frac{G_2}{45}[8 K^{(1)}_{isjrkqmp}-7 K^{(2)}_{isjrkmpq}-K^{(3)}_{isjrkpmq}-7 K^{(4)}_{isjpkrmq}+5 K^{(5)}_{isjqkmpr}\nonumber \\&+11 K^{(6)}_{irjskpmq}-4 K^{(7)}_{irjskmpq}+ 5 K^{(8)}_{imjkpsqr}]+\frac{G_3}{20}[-18 K^{(1)}_{isjrkqmp}+17 K^{(2)}_{isjrkmpq}\nonumber\\
&+K^{(3)}_{isjrkpmq}+27 K^{(4)}_{isjpkrmq}-15 K^{(5)}_{isjqkmpr}-36 K^{(6)}_{irjskpmq}+9 K^{(7)}_{irjskmpq}], \label{eq:fourthmomentfinal}
\end{align}
where the $R_\lambda$-dependent values of $G_1$, $G_2$ and $G_3$ are again obtained from DNS.

\subsection{Fourth-moment of the spheroid orientation distribution}\label{sec:fourthmoment}
Having derived the form for the fourth moment of the velocity-gradient tensor in homogeneous isotropic turbulence, we proceed to the derivation of $\langle(1-p_3)^2\rangle$(equation (5) in the main text). In the rapid-settling limit, 
\begin{align}
\langle(1-p_3)^2\rangle\approx \dfrac{1}{4}\langle (p_1^2+p_2^2)^2\rangle,
\end{align}
which involves the quantities $\langle p_1^4\rangle$, $\langle p_2^4\rangle$ and $\langle p_1^2p_2^2\rangle$. We raise equations (\ref{eq:p1}) and (\ref{eq:p2}) to the fourth power and also multiply the squares of the two and ensemble average the resulting terms; isotropy arguments are then used to rule out averages of the form $\langle\bm{EEES}\rangle$, for instance. This finally leads to the following expressions:
\begin{subequations} \label{eq:pfourth}
\begin{align} 
  \langle p_1^4 \rangle &\approx (\dfrac{8\pi Y_A Y_c}{f_I(\kappa) X_A})^4[\langle S_{31}S_{31}S_{31}S_{31}\rangle+6B^2\langle S_{31}S_{31}E_{31}E_{31}\rangle\nonumber\\&+B^4\langle E_{31}E_{31}E_{31}E_{31}\rangle]\dfrac{1}{Fr_\eta^8},\\
  \langle p_2^4 \rangle &\approx (\dfrac{8\pi Y_A Y_c}{f_I(\kappa) X_A})^4[\langle S_{32}S_{32}S_{32}S_{32}\rangle+6B^2\langle S_{32}S_{32}E_{32}E_{32}\rangle\nonumber\\&+B^4\langle E_{32}E_{32}E_{32}E_{32}\rangle]\dfrac{1}{Fr_\eta^8},\\
  \langle p_1^2p_2^2 \rangle &\approx (\dfrac{8\pi Y_A Y_c}{f_I(\kappa) X_A})^4[\langle S_{31}S_{31}S_{32}S_{32}\rangle+B^2(\langle S_{31}S_{31}E_{32}E_{32}\rangle\nonumber\\&+\langle S_{32}S_{32}E_{31}E_{31}\rangle +4\langle S_{31}S_{32}E_{31}E_{32}\rangle)+B^4\langle E_{31}E_{31}E_{32}E_{32}\rangle]\dfrac{1}{Fr_\eta^8},
\end{align}
\end{subequations}
keeping in mind that $p_{1,2}\ll p_3\approx 1$.
To calculate the averages occuring in~(\ref{eq:pfourth}), for instance, $\langle S_{31}S_{31}S_{31}S_{31}\rangle$, we use $\bm{S}=\frac{\bm{\Gamma}-\bm{\Gamma}^\dagger}{2}$ and $\bm{E}=\frac{\bm{\Gamma}+\bm{\Gamma}^\dagger}{2}$. This particular ensemble average then takes the form:
\begin{align}
 \langle S_{31}S_{31}S_{31}S_{31}\rangle &= \dfrac{1}{16}(\langle \Gamma_{31}\Gamma_{31}\Gamma_{31}\Gamma_{31}\rangle-\langle \Gamma_{31}\Gamma_{31}\Gamma_{31}\Gamma_{13}\rangle-\langle \Gamma_{31}\Gamma_{31}\Gamma_{13}\Gamma_{31}\rangle\nonumber \\&+\langle \Gamma_{31}\Gamma_{31}\Gamma_{13}\Gamma_{13}\rangle-\langle \Gamma_{31}\Gamma_{13}\Gamma_{31}\Gamma_{31}\rangle+\langle \Gamma_{31}\Gamma_{13}\Gamma_{31}\Gamma_{13}\rangle\nonumber \\&+\langle \Gamma_{31}\Gamma_{13}\Gamma_{13}\Gamma_{31}\rangle-\langle \Gamma_{31}\Gamma_{13}\Gamma_{13}\Gamma_{13}\rangle-\langle \Gamma_{13}\Gamma_{31}\Gamma_{31}\Gamma_{31}\rangle\nonumber \\&+\langle \Gamma_{13}\Gamma_{31}\Gamma_{31}\Gamma_{13}\rangle+\langle \Gamma_{13}\Gamma_{31}\Gamma_{13}\Gamma_{31}\rangle-\langle \Gamma_{13}\Gamma_{31}\Gamma_{13}\Gamma_{13}\rangle\nonumber \\&+\langle \Gamma_{13}\Gamma_{13}\Gamma_{31}\Gamma_{31}\rangle-\langle \Gamma_{13}\Gamma_{13}\Gamma_{31}\Gamma_{13}\rangle-\langle \Gamma_{13}\Gamma_{13}\Gamma_{13}\Gamma_{31}\rangle\nonumber \\&+\langle \Gamma_{13}\Gamma_{13}\Gamma_{13}\Gamma_{13}\rangle).
\end{align}
Using ~(\ref{eq:fourthmomentfinal}) for the fourth moment, the terms occuring on the right-hand side above can be evaluated. Adopting the same procedure to calculate all the ensemble averages in~(\ref{eq:pfourth}), we get:
\begin{subequations}
\begin{align}
 \langle p_1^4 \rangle &\approx (\dfrac{8\pi Y_A Y_c}{f_I(\kappa) X_A})^4[\frac{1}{80}(45 G_1+64 G_2-324 G_3)+\frac{B^2}{40}(-45 G_1 + 8 G_2 + 162 G_3)\nonumber\\&+\dfrac{9G_1B^4}{16}]\dfrac{1}{Fr_\eta^8},\\
  \langle p_2^4 \rangle &\approx (\dfrac{8\pi Y_A Y_c}{f_I(\kappa) X_A})^4[\frac{1}{80}(45 G_1+64 G_2-324 G_3)+\frac{B^2}{40}(-45 G_1 + 8 G_2 + 162 G_3)\nonumber\\&+\dfrac{9G_1B^4}{16}]\dfrac{1}{Fr_\eta^8},\\
  \langle p_1^2 p_2^2 \rangle &\approx (\dfrac{8\pi Y_A Y_c}{f_I(\kappa) X_A})^4[\dfrac{1}{240}(45 G_1+64G_2-324G_3)+\dfrac{B^2}{120}(-45 G_1+8G_2+162G_3)\nonumber \\&+\dfrac{3G_1B^4}{16}]\dfrac{1}{Fr_\eta^8}.
\end{align}
\end{subequations}
Hence, we obtain the following expression for the fourth moment of the orientation distribution for an oblate spheroid: 
\begin{align}
\langle (1\!-\!p_3)^2 \rangle\! \approx \frac{1}{4}\langle (1\!-\!p_3^2)^2 \rangle\! \!&\approx& \!\! \dfrac{1}{4}(\!\dfrac{8\pi Y_A Y_c}{f_I(\kappa) X_A}\!)^4 [M_1\!+\!B^2 M_2\! +\! B^4 M_3]\frac{1}{Fr_\eta^8}, \label{eq:fourthmomentorient}
\end{align}
where $M_1=\dfrac{3G_1}{2}+\dfrac{32G_2}{15}-\dfrac{162G_3}{15}$, $M_2=-3G_1+\dfrac{8G_2}{15}+\dfrac{54G_3}{5}$ and $M_3=\dfrac{3 G_1}{2}$.
\\
\\
\\
Similarly, the fourth moment for a nearly broadside-on prolate spheroid($p_3 \ll p_{1,2} \approx 1$) is given by:
\begin{align}
  \langle p_3^4 \rangle\! &\approx \!\! (\!\dfrac{8\pi Y_A Y_c}{f_I(\kappa) X_A}\!)^4 [M_1\!+\!B^2 M_2\! +\! B^4 M_3]\frac{1}{Fr_\eta^8}, \label{eq:fourthmomentfinalpro}
\end{align}
where $M_1=\dfrac{9G_1}{16}+\dfrac{4G_2}{5}-\dfrac{81G_3}{20}$, $M_2=-\dfrac{9G_1}{8}+\dfrac{G_2}{5}+\dfrac{81G_3}{20}$ and $M_3=\dfrac{9 G_1}{16}$.
\section{Parameter space for the direct numerical simulations(DNS)}
\begin{table}
\small
\centering
\begin{tabular}{ |c|c|c|c|c|c|c|c| } 
 \hline
  Parameters in \cite{pumir_18} & $\kappa$ & $Fr_\eta=U_s/u_\eta$ & $St_\eta$ & $Re_s$ & $Re_{\dot{\gamma}}$ & $L(\mu m)$ & $L/\eta$\\
 \hline \hline
 
 \multirow{3}{*}{$R_\lambda=200$}  & 0.05 & 4.2 & 0.4 & 0.8 & 0.036 & 107 & 0.22 \\ 
 
  & 0.05 & 3.1 & 0.29 & 0.5 & 0.026 & 92 & 0.18\\ 
 
  & 0.05 & 1.05 & 0.1 & 0.1 & 0.01 & 54 & 0.1\\ 
  \cline{2-8}
  \multirow{3}{*}{$\epsilon=0.51 m^2s^{-3}$} & 0.02 & 3.1 & 0.296 & 0.8 & 0.066 & 145 & 0.29 \\ 
  
  & 0.02 & 2.3 & 0.216 & 0.5 & 0.048 & 124 & 0.25\\ 
 
  & 0.02 & 0.78 & 0.074 & 0.1 & 0.016 & 73 & 0.15\\ 
  \cline{2-8}
  \multirow{3}{*}{$\nu=1.1\times 10^{-5} m^2 s^{-1}$} & 0.01 & 2.5 & 0.24 & 0.8 & 0.1 & 183 & 0.37 \\ 
 
  & 0.01 & 1.8 & 0.17 & 0.5 & 0.076 & 157 & 0.32\\ 
 
  & 0.01 & 0.6 & 0.06 & 0.1 & 0.026 & 92 & 0.18\\ 
 \hline 
 
 \multirow{3}{*}{$R_\lambda=150$}  & 0.05 & 3.7 & 0.4 & 0.8 & 0.04 & 107 & 0.22 \\ 
 
  & 0.05 & 2.7 & 0.29 & 0.5 & 0.03 & 92 & 0.18\\ 
 
  & 0.05 & 0.9 & 0.1 & 0.1 & 0.01 & 54 & 0.1\\ 
  \cline{2-8}
  \multirow{3}{*}{$\epsilon=0.0246 m^2s^{-3}$} & 0.02 & 2.7 & 0.296 & 0.8 & 0.08 & 145 & 0.29 \\ 
  
  & 0.02 & 2.0 & 0.216 & 0.5 & 0.06 & 124 & 0.25\\ 
 
  & 0.02 & 0.7 & 0.074 & 0.1 & 0.02 & 73 & 0.15\\ 
  \cline{2-8}
  \multirow{3}{*}{$\nu=1.1\times 10^{-5} m^2 s^{-1}$} & 0.01 & 2.1 & 0.24 & 0.8 & 0.14 & 183 & 0.37 \\ 
 
  & 0.01 & 1.6 & 0.17 & 0.5 & 0.1 & 157 & 0.32\\ 
 
  & 0.01 & 0.5 & 0.06 & 0.1 & 0.03 & 92 & 0.18\\ 
 \hline
 
  \multirow{3}{*}{$R_\lambda=96$}  & 0.05 & 7.4 & 0.1 & 0.8 & 0.01 & 107 & 0.1 \\ 
 
  & 0.05 & 5.4 & 0.074 & 0.5 & 0.008 & 92 & 0.09\\ 
 
  & 0.05 & 1.8 & 0.025 & 0.1 & 0.003 & 54 & 0.05\\ 
  \cline{2-8}
  \multirow{3}{*}{$\epsilon=0.001562 m^2s^{-3}$} & 0.02 & 5.4 & 0.075 & 0.8 & 0.02 & 145 & 0.15 \\ 
  
  & 0.02 & 4.0 & 0.05 & 0.5 & 0.016 & 124 & 0.13\\ 
 
  & 0.02 & 1.4 & 0.019 & 0.1 & 0.005 & 73 & 0.07\\ 
  \cline{2-8}
  \multirow{3}{*}{$\nu=1.1\times 10^{-5} m^2 s^{-1}$} & 0.01 & 4.3 & 0.06 & 0.8 & 0.03 & 183 & 0.19 \\ 
 
  & 0.01 & 3.2 & 0.04 & 0.5 & 0.02 & 157 & 0.16\\ 
 
  & 0.01 & 1.1 & 0.015 & 0.1 & 0.008 & 92 & 0.09\\ 
 \hline 
 
  \multirow{3}{*}{$R_\lambda=47$}  & 0.05 & 13.4 & 0.025 & 0.8 & 0.00354 & 107 & 0.05 \\ 
 
  & 0.05 & 9.8 & 0.019 & 0.5 & 0.0026 & 92 &0.04\\ 
 
  & 0.05 & 3.4 & 0.006 & 0.1 & 0.00088 & 54 & 0.03\\ 
  \cline{2-8}
  \multirow{3}{*}{$\epsilon=9.8\times10^{-5} m^2s^{-3}$} & 0.02 & 10 & 0.019 & 0.8 & 0.006 & 145 & 0.07 \\ 
  
  & 0.02 & 7.3 & 0.014 & 0.5 & 0.0047 & 124 & 0.06\\ 
 
  & 0.02 & 2.5 & 0.0047 & 0.1 & 0.0016 & 73 & 0.04\\ 
  \cline{2-8}
  \multirow{3}{*}{$\nu=1.1\times 10^{-5} m^2 s^{-1}$} & 0.01 & 7.9 & 0.015 & 0.8 & 0.01 & 183 & 0.09 \\ 
 
  & 0.01 & 5.8 & 0.011 & 0.5 & 0.0075 & 157 & 0.08\\ 
 
  & 0.01 & 2 & 0.0037 & 0.1 & 0.0026 & 92 & 0.05\\ 
  \cline{2-8}
  & 0.1 & 17 & 0.032 & 0.8 & 0.0022 & 85 & 0.043 \\ 
 
  & 0.1 & 12.4 & 0.023 & 0.5 & 0.0016 & 73 & 0.037\\ 
 
  & 0.1 & 4.2 & 0.008 & 0.1 & 0.00056 & 43 & 0.02\\ 
 \hline
 
\end{tabular}
\caption{Parameters for DNS runs in Figures 1,2 and 4 in the main manuscript}
\label{tab:Fig124}
\end{table}

\begin{table}
\small
\centering
\begin{tabular}{ |c|c|c|c|c|c|c|c| } 
 \hline
  Parameters in \cite{pumir_18} & $\kappa$ & $Fr_\eta=U_s/u_\eta$ & $St_\eta$ & $Re_s$ & $Re_{\dot{\gamma}}$ & $L(\mu m)$ & $L/\eta$\\
 \hline \hline
 
 \multirow{2}{*}{$R_\lambda=150$} & 0.999 & 5.5 & 0.64 & 0.36 & 0.004 & 33 & 0.07 \\ 
 
  & 0.91 & 5.1 & 0.6 & 0.34 & 0.004 & 33 & 0.07 \\  
 
 \multirow{2}{*}{$\epsilon=0.0246 m^2s^{-3}$} & 0.67 & 3.9 & 0.46 & 0.26 & 0.004 & 33 & 0.07 \\  
  
  & 0.53 & 3.2 & 0.37 & 0.21 & 0.004 & 33 & 0.07 \\
  
  \multirow{2}{*}{$\nu=1.1\times 10^{-5} m^2 s^{-1}$} & 0.1 & 0.64 & 0.07 & 0.04 & 0.004 & 33 & 0.07 \\
  
  & 0.05 & 0.32 & 0.04 & 0.02 & 0.004 & 33 & 0.07 \\
   
 \hline
 
 \multirow{2}{*}{$R_\lambda=96$} & 0.999 & 11.9 & 0.16 & 0.4 & 0.001 & 33 & 0.03 \\ 
 
  & 0.91 & 11 & 0.15 & 0.37 & 0.001 & 33 & 0.03 \\  
 
  \multirow{2}{*}{$\epsilon=0.001562 m^2s^{-3}$} & 0.67 & 8.5 & 0.11 & 0.28 & 0.001 & 33 & 0.03 \\  
  
  & 0.53 & 6.9 & 0.09 & 0.23 & 0.001 & 33 & 0.03 \\
  
  \multirow{2}{*}{$\nu=1.1\times 10^{-5} m^2 s^{-1}$} & 0.1 & 1.4 & 0.019 & 0.05 & 0.001 & 33 & 0.03 \\
  
  & 0.05 & 0.7 & 0.0095 & 0.02 & 0.001 & 33 & 0.03 \\
   
 \hline
 
  \multirow{2}{*}{$R_\lambda=47$} & 0.999 & 31.6 & 0.06 & 0.63 & 0.0004 & 39 & 0.02 \\ 
 
  & 0.91 & 29.3 & 0.05 & 0.59 & 0.0004 & 39 & 0.02 \\  
 
 \multirow{2}{*}{$\epsilon=9.8\times 10^{-5} m^2s^{-3}$} & 0.67 & 22.5 & 0.04 & 0.45 & 0.0004 & 39 & 0.02 \\  
  
  & 0.53 & 18.3 & 0.03 & 0.36 & 0.0004 & 39 & 0.02 \\
  
 \multirow{2}{*}{$\nu=1.1\times 10^{-5} m^2 s^{-1}$} & 0.1 & 3.7 & 0.007 & 0.07 & 0.0004 & 39 & 0.02 \\
  
  & 0.05 & 1.9 & 0.003 & 0.04 & 0.0004 & 39 & 0.02 \\
   
 \hline
 
\end{tabular}
\caption{Parameters for DNS runs in Figure 3 in the main manuscript}
\label{tab:Fig3}
\end{table}
Figures 1, 2 and 4 in the main manuscript showcase DNS data and RST results based on the parameters listed in table \ref{tab:Fig124} while figure 3 in the main manuscript is based on table \ref{tab:Fig3}. The physical parameters listed in the table above have been drawn from \cite{pumir_18}. Except for $R_\lambda$, the other parameters are representative of the atmospheric scenario such as ice-crystal size and aspect ratio (\cite{klett_book}, \cite{Veal_1970}) and the dissipation rates(\cite{siewert_14}). The $R_\lambda$'s are lower than in the atmospheric case and this leads to less intermittent distributions.

\section{Results}
\subsection{Preferential Concentration}
A suspension of spherical particles in a turbulent flow is no longer spatially homogeneous when $Fr_\eta, St_\eta \sim \mathcal{O}(1)$\,\cite{BecRay_2014,eaton_Review}.  Preferential sampling of regions of low vorticity by inertial particles, together with a
sweeping effect in presence of gravity, leads to
enhanced settling velocities\,\cite{maxey_1987,maxey_1993,collins_2014}. Fig. 3 in the manuscript 
shows this to be true for the suspensions of spheroids considered here.
In this figure, it is seen that, for large $Fr_\eta$, the time-averaged settling speeds from the DNS agree with
the orientational averages for $R_\lambda = 47$ and $96$\,(the $Fr_\eta$ required for this agreement increases with increasing $St_\eta$). For finite $Fr_\eta$ and
$St_\eta$, however, the time averages consistently exceed the orientation averaged
estimates.
\begin{figure}[htbp]
\includegraphics[width=0.49\columnwidth]{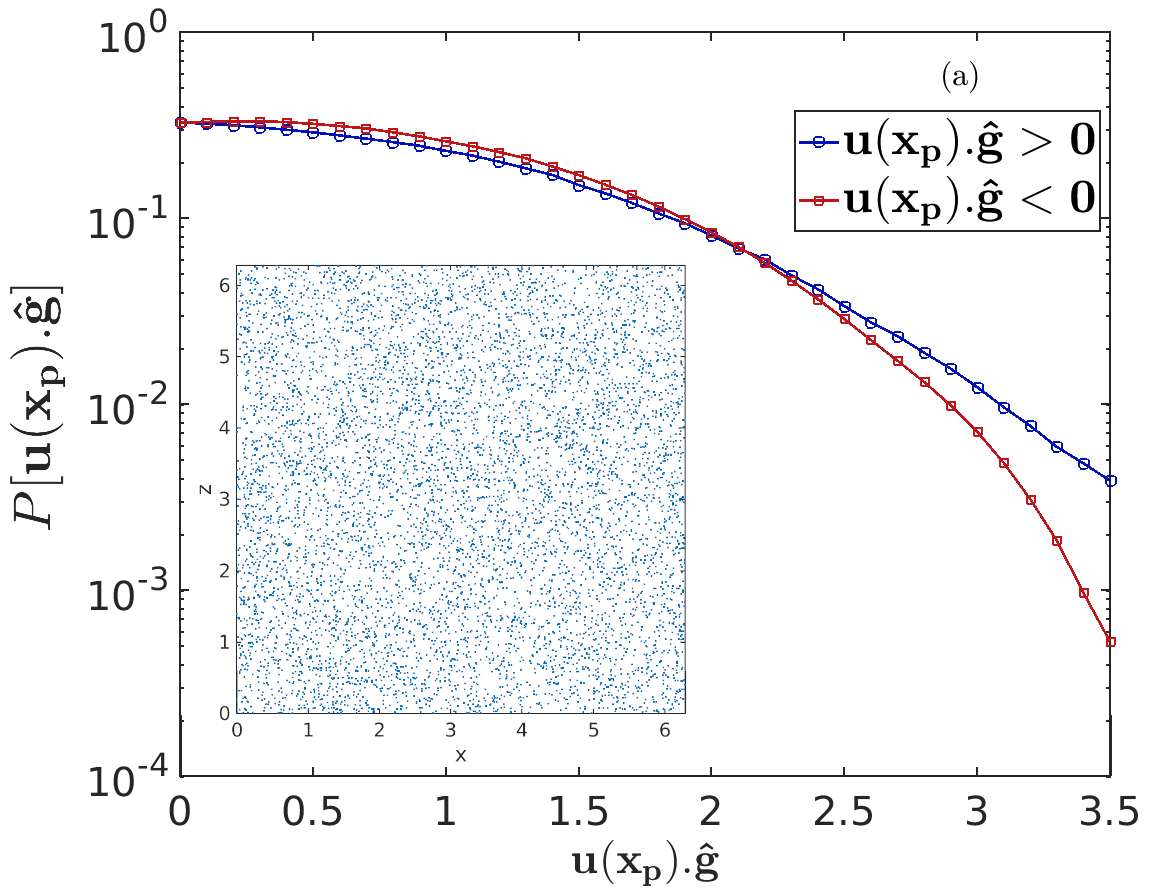}
\includegraphics[width=0.49\columnwidth]{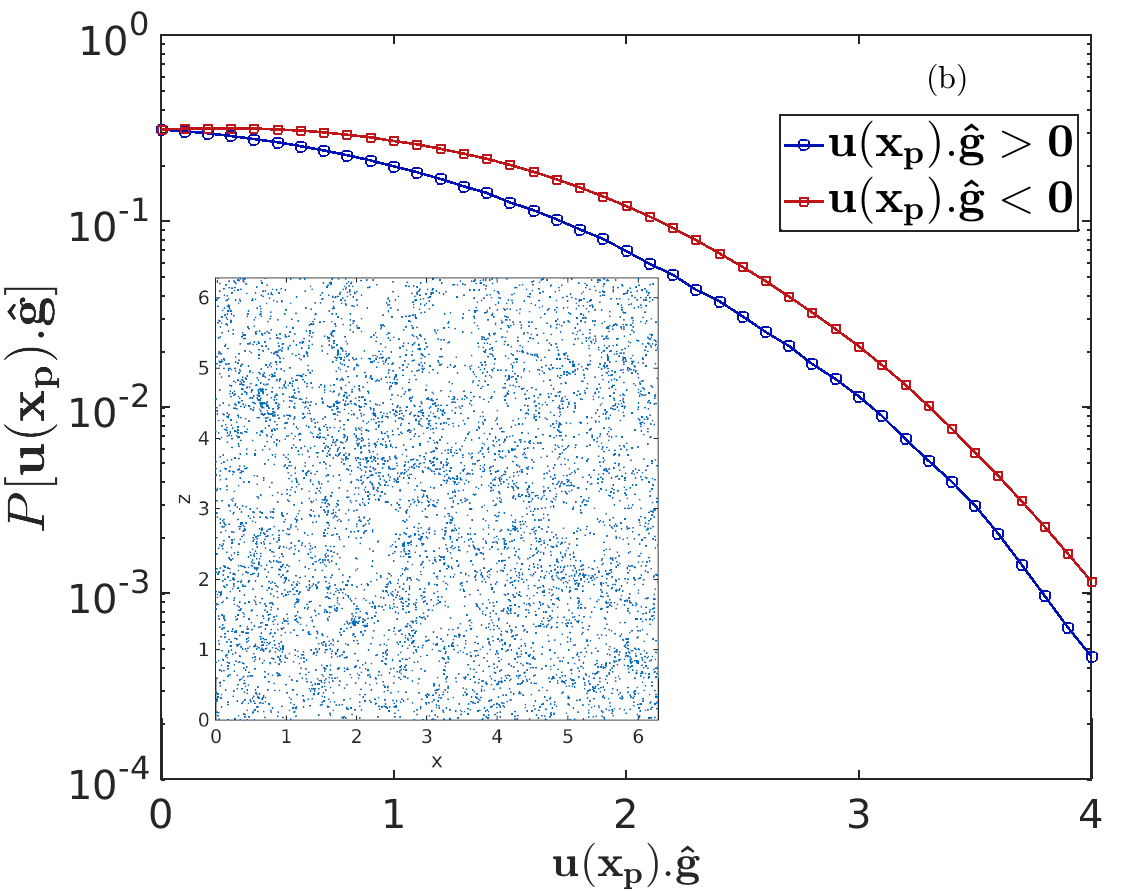}
\caption{Distributions of vertical velocity along a particle trajectory, conditioned on being parallel or anti-parallel to gravity (see legend) for (a)$St_\eta\!=\!0.025$, $Fr_\eta\!=\!13.45$, $R_\lambda\!=\!47$, (b)$St_\eta\!=\!0.4$, $Fr_\eta\!=\!3.66$, $R_\lambda\!=\!150$. The insets show corresponding snapshots of the particle locations in a slice of height and width 2$\pi$ each with thickness=$50 l_\eta$ in (a) and $10l_\eta$ in (b). } 
\label{fig:prefconc}
\end{figure}

Fig.~\ref{fig:prefconc} above confirms that the discrepancy between the time and
orientation-averaged settling speeds in Fig. 3 (in the main manuscript) is
due to the preferential sweeping effect. The insets show
instantaneous snapshots of particle positions for (a) $R_\lambda = 47$ and (b) 
$150$. The particle concentration field remains spatially homogeneous for
$R_\lambda = 47$, in which case $St_\eta=0.025$; while there is clear evidence of clustering for $R_\lambda =
150$. The spatial inhomogeneity in the particle concentration fields has also been characterized via pair-distribution functions (not shown). The probability distributions for the occurrence of upflow\,($u_3 > 0$)
and downflow\,($u_3 < 0$) along particle trajectories have been shown alongside in Fig.\ref{fig:prefconc}. The
enhanced sampling of downflow regions for $R_\lambda = 150$ is evidence of
preferential sweeping. Preferential sweeping effects in the anisotropic particle suspensions examined here should not come as a surprise since the particle orientation distributions are
localized around the broadside-on orientation, and the variation of settling velocity with orientation is therefore minimal, implying a resemblance
to spherical particles.


\begin{thebibliography}{60}
\bibitem{footnote3} The two points in Fig.~\ref{fig:moments}(b) which, despite having the two highest $Fr_\eta$, have poor agreement between the analytical and DNS values, correspond to the two shortest tails in Fig.~\ref{fig:pdf}(b). The simulation needs to run for longer times to capture these tails.
\end{thebibliography}
\end{document}